\documentclass[fleqn,usenatbib]{mnras}
\usepackage{newtxtext,newtxmath}

\usepackage[T1]{fontenc}
\usepackage{ae,aecompl}

\usepackage{graphicx}	
\usepackage{amsmath}	
\usepackage{nccmath}

\title[Multiple shock fronts in A1914]{On the detection of multiple shock fronts in A1914 using deep \textit{Chandra} X-ray observations}
\author[Rahaman et al.]{
Majidul Rahaman,$^{1}$\thanks{E-mail: phd1601121007@iiti.ac.in}
Ramij Raja,$^{1}$
and Abhirup Datta$^{1}$
\\
$^{1}$\href{http://astronomy.iiti.ac.in/index.html}{Discipline of Astronomy, Astrophysics and Space Engineering}, \href{http://www.iiti.ac.in/}{Indian Institute Of Technology Indore}, Indore, India\\
}

\date{Accepted 2021 October 22. Received 2021 October 21; in original form 2021 April 14.}

\pubyear{2021}
\begin{document}
\label{firstpage}
\pagerange{\pageref{firstpage}--\pageref{lastpage}}
\maketitle

\begin{abstract}
Here, we report the new detection of three shock fronts using archival \textit{Chandra} X-ray observations in Abell 1914, which also hosts a radio halo, a radio phoenix, and a head-tail galaxy. In this study, we report the X-ray shock front at the position of the radio phoenix, which further strengthens the scenario that radio phoenix traces old plasma that gets lit up when compressed by shock passage. We further analyze the thermodynamic structure of the cluster in detail. We create temperature maps of A1914 using three different techniques, Adaptive Circular Binning (ACB), Weighted Voronoi Tessellations (WVT), and Contour binning (Contbin) method. These thermodynamic maps, along with the pseudo pressure and the pseudo entropy maps for the cluster, are evidence of disturbed morphology produced by multiple merger events. These merger events create cluster-wide turbulence, which may re-accelerate the relativistic particles and result in a radio halo within the cluster. Further, comparing X-ray and radio images reveals that the radio halo is contained within two X-ray shock fronts.
Our analysis suggests that A1914 has both equatorial shock and axial shock within the cluster's ICM. We proposed a dual merging scenario based on the shock position and analysis of the thermodynamic maps obtained from the deeper \textit{Chandra} X-ray observations.

\end{abstract}


\section{Introduction}
Galaxy clusters grow via major and minor mergers, the most energetic events in the Universe after the Big Bang. When cluster merges, shock and cold fronts are triggered into the intracluster medium (ICM). 
Major mergers release as much as $10^{+64}$ erg of energy on Gyr timescales.
Most of the gravitational energy released during cluster merger events is converted into thermal energy via low-Mach number shocks (M $\le$3) and turbulence in the ICM  \citep{Ryu_2003ApJ...593..599R}.
Shocks and turbulence generated by the merger would heat the intracluster gas and accelerate ultra-relativistic particles and amplify magnetic fields that coexist with the thermal plasma.
These ultra-relativistic electrons within the thermal plasma unveil themselves through synchrotron radio emission in the shape of radio halos, relics, and radio phoenixes \citep{Kempner2004rcfg.proc..335K,Brunetti2014IJMPD..2330007B,van_Weeren2019SSRv..215...16V}.
While these components' energy density is a small fraction of the thermal gas pressure, they can substantially alter the physics of the ICM.
The origin of the radio halo is debated between two models: hadronic \citep{Dennison1980ApJ...239L..93D,Blasi1999APh....12..169B,Dolag2000A&A...362..151D,Miniati2001ApJ...559...59M,Miniati2001ApJ...562..233M,Pfrommer2008MNRAS.385.1211P,Enblin2011A&A...527A..99E} and turbulent re-acceleration \citep{Brunetti2001MNRAS.320..365B,Petrosian2001ApJ...557..560P,Donnert2013MNRAS.429.3564D}. In the hadronic model, radio-emitting electrons are produced in the hadronic interaction between CR protons and ICM protons. However, in the re-acceleration model, a population of seed electrons is re-accelerated during powerful ICM turbulence states by cluster merger.
Radio halos are predominately found in merging clusters \citep{cassano2013ApJ...777..141C,kale2015A&A...579A..92K,Cuciti2015A&A...580A..97C}.
Elongated shaped cluster radio shocks/relics are mostly found in the outskirts of the cluster and trace shock waves \citep{Feretti2012A&ARv..20...54F,van_Weeren2019SSRv..215...16V}. Radio phoenixes are the less studied object found in the ICM, and their origin is still not clear \citep{van_Weeren2019SSRv..215...16V}. The current favored scenario is that phoenixes are a manifestation of fossil plasma in galaxy clusters from past AGN activity \citep{van_Weeren2019SSRv..215...16V}.
When merger shocks compress fossil plasma, the momentum of relativistic electrons and magnetic field strength increased. This compression of fossil plasma results in synchrotron radiation, which is characterized by ultra-steep or curved radio spectrum  \citep{Enblin2001A&A...366...26E}
 Diffuse radio emission produced by the above process has complex and filamentary morphology \citep{Enblin2002MNRAS.331.1011E}. However, there was no observational evidence of the shock near the radio phoenix. Therefore, the study of the origin of radio phoenix is essential.

Here, we make the radio–X-ray comparison for A1914, a merging cluster at redshift (z) = 0.168.
It is a merging cluster with complex geometry \citep{Barrena2013MNRAS.430.3453B,Mann2012MNRAS.420.2120M,Botteon_2018MNRAS.476.5591B}.
Irregular mass distribution was reported by \citet{Okabe2008PASJ...60..345O} using weak lensing data. Previous Chandra studies highlighted a heated ICM with temperature peak in the cluster center \citep{Govoni_2004ApJ...605..695G,Baldi2007ApJ...666..835B,Botteon_2018MNRAS.476.5591B}.
The presence of a radio halo in A1914 was suggested by \citet{Giovannini_1999NewA....4..141G} from an NVSS search. The radio halo was detected by \citet{Kempner2001ApJ...548..639K} in the Westerbork Northern Sky Survey (WENSS) at 0.3 GHz and confirmed by \citet{Bacchi_2003A&A...400..465B} in deep VLA observations at 1.4 GHz.
In \citet{Mandal_a2019A&A...622A..22M}, they resolve the diffuse radio structure properly using high-resolution LOFAR radio observations. They found that the overall radio emission of A1914 is unlikely to be a single radio halo but rather the superposition of a revived fossil plasma source (radio phoenix), a radio halo, and a head-tail radio galaxy.

To investigate the origin of the radio emissionin A1914, we study A1914 in detail using archival \textit{Chandra} X-ray observations. The paper is organized as follows. We present information about data analysis in section \ref{sec:X-ray}. We present our results on the cluster structure in section \ref{results}.  In sections \ref{Discussions} , we discussed our results. In section \ref{Conclusions}, we conclude our findings.
Unless otherwise stated, errors are presented at the 68 percent confidence level. 

Throughout the paper we assume a $\Lambda$CDM cosmology with $H_0 = 70$ km s$^{-1}$ Mpc$^{-1}$, $\Omega_{m} = 0.3$ and $\Omega_\Lambda = 0.7$.
Within this cosmology, 1 arcmin corresponds to 0.18 Mpc at the cluster redshift (z = 0.168).

\begin{table}
	\caption{Details of \textit{Chandra} X-ray archival observations.}
	\label{tab:data_table}
	\begin{tabular}{lccr}
		\hline
 		ObsId's & Obs. Cycle  & Clean \\
        &   & Exposure(ks) \\
		\hline
		542 & 01  & 8 \\
		3593 & 04 & 18 \\
		18252, 20023, 20024  &  17   & 120 \\
		20025, 20026 &     & \\
        \hline
        Total & & 146
	\end{tabular}
\end{table}

\begin{figure*}
\centering
\begin{tabular}{lccr}
\includegraphics[width=\columnwidth]{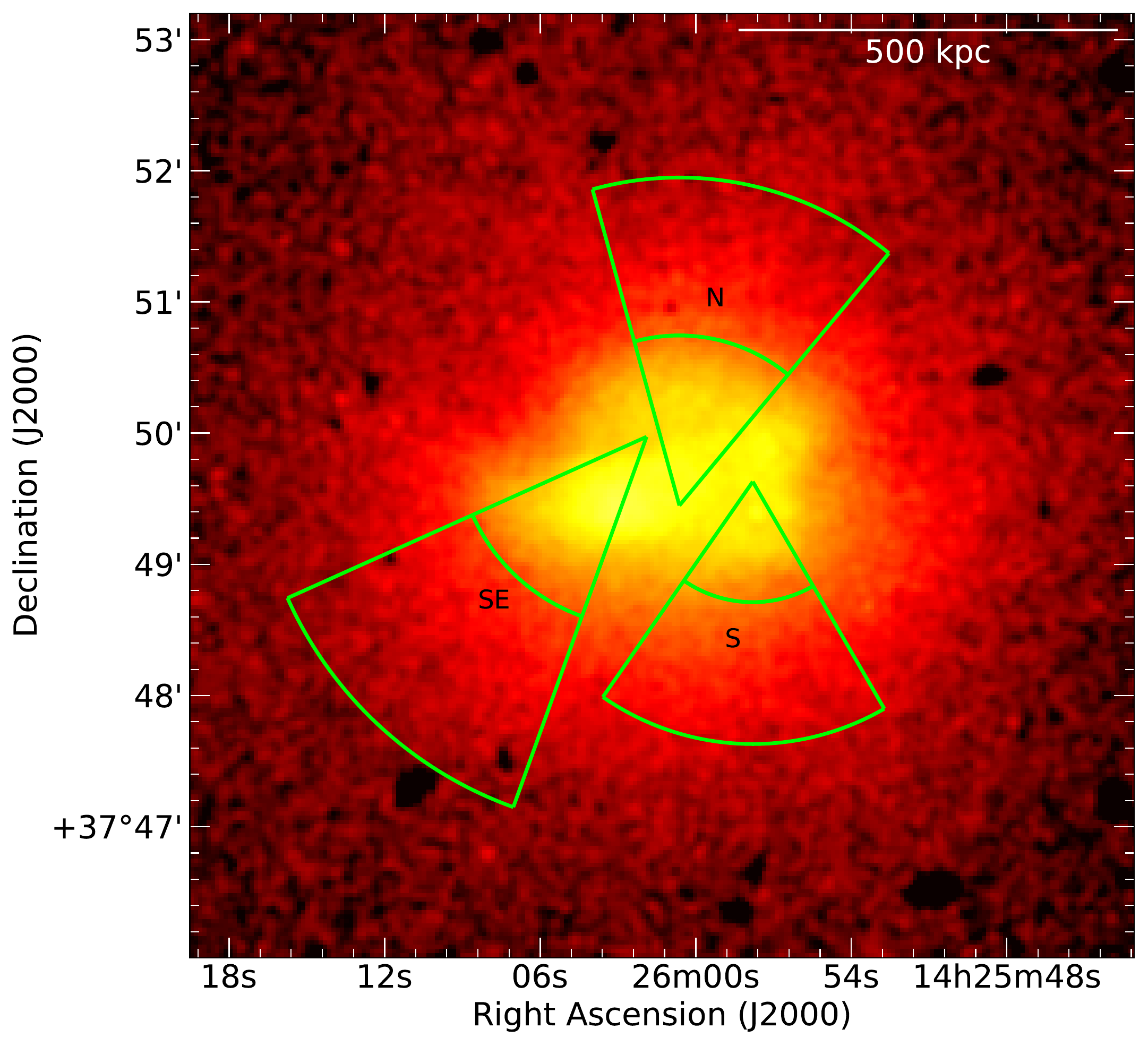} &
\includegraphics[width=\columnwidth]{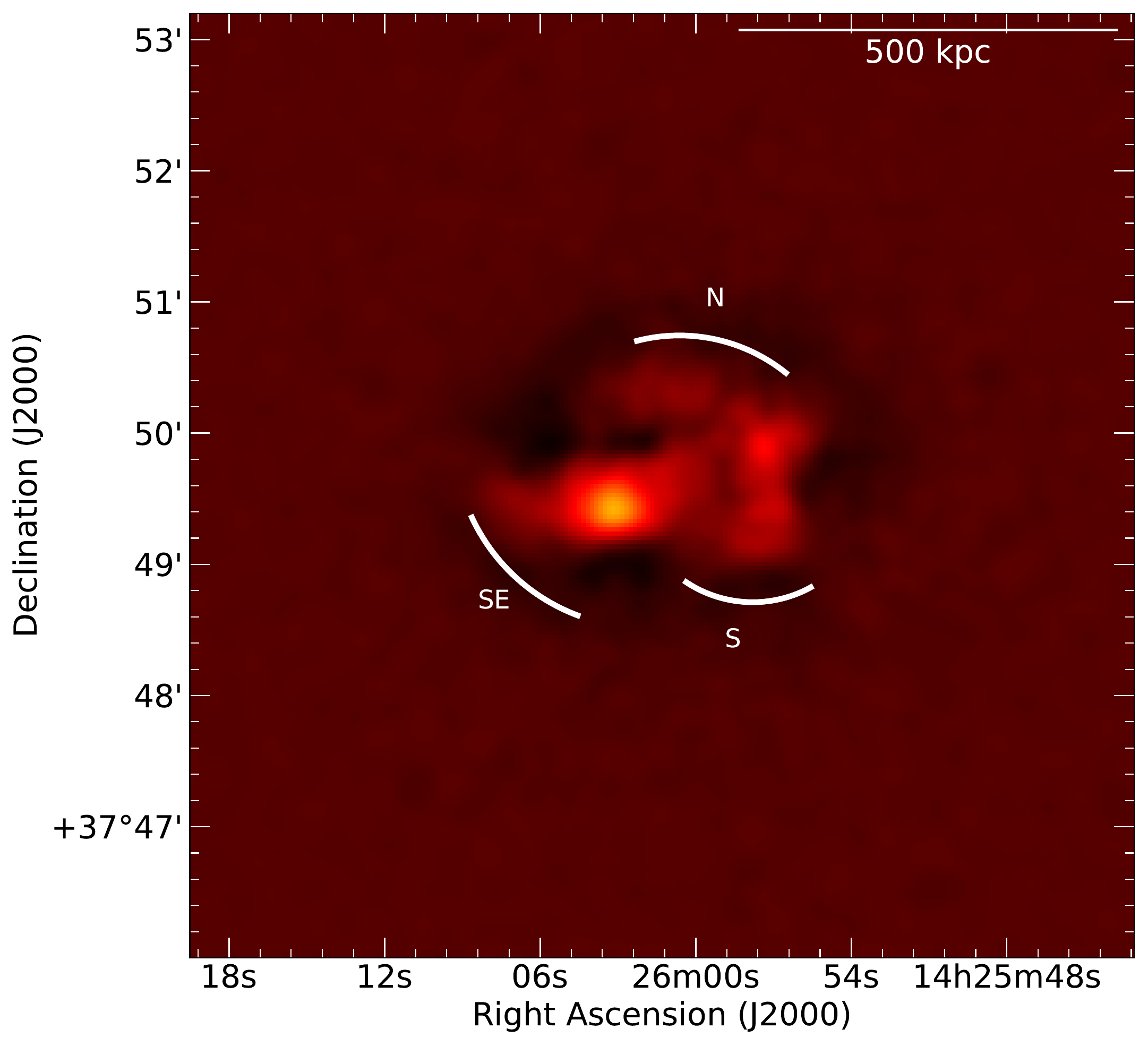} \\
\includegraphics[width=\columnwidth,height=7.8cm]{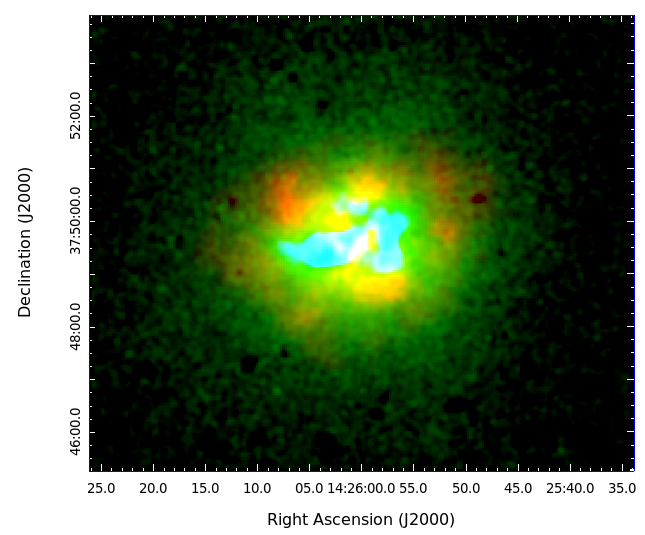} &
\includegraphics[width=\columnwidth]{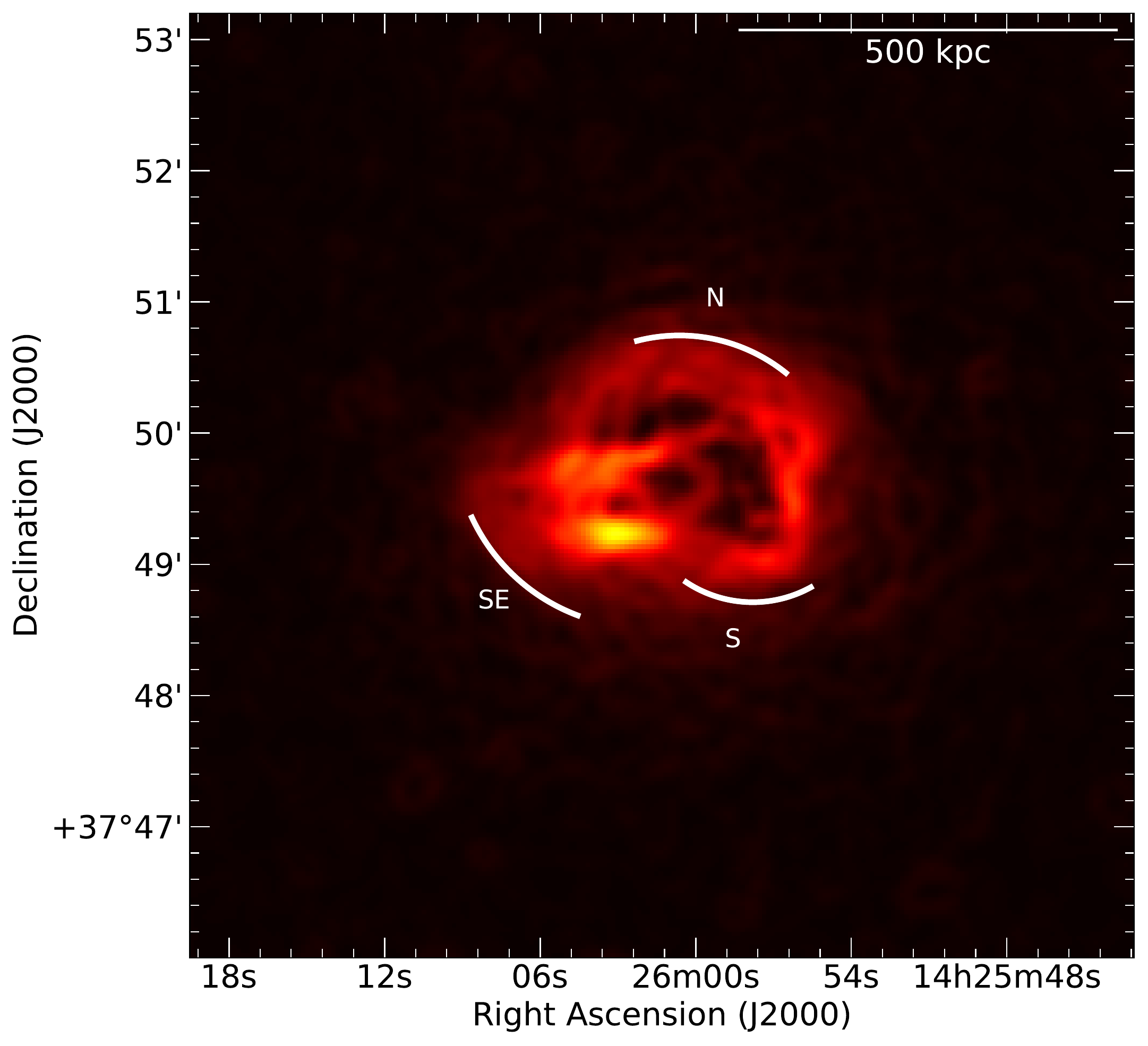} \\
\end{tabular}
\caption{Top left: Exposure corrected background subtracted, point source removed surface brightness map of A1914 cluster. Wedge regions (green) represent the newly detected X-ray shock wave positions.
Top right: The 0.7-8.0 keV unsharp-masked \textit{Chandra} X-ray surface brightness image was derived by subtracting a 10$\sigma$ wide Gaussian kernel-smoothed image from that smoothed with a 2$\sigma$ wide Gaussian kernel. Arcs (N, S, and SE) (white) represent the positions of shock waves.
Bottom Left: The RGB image of the \textit{Chandra} X-ray surface brightness map of A1914, where colors are represents as red 2.0-7.0 keV, green 1.2-2.0 keV, and blue 0.7-1.2 keV X-ray surface brightness. In this tricolor image, some substructure can be visible, which resembles the substructure found in the \textit{Chandra} X-ray temperature maps (Figure \ref{fig:ACB_Tmap}).
Bottom Right: Gaussian gradient magnitude (GGM) filtered images on scales of 4. Arcs (N, S, and SE) (white) represent the positions of shock waves.
}
\label{fig:X-Ray-SB}
\end{figure*}

\section{X-ray observations}  \label{sec:X-ray}
We analyzed \textit{Chandra} archival X-ray observations of A1914, which was observed with ACIS-I over series of programs, listed in Table \ref{tab:data_table}. All the data were observed in VFAINT mode. In total, the effective exposure of 146 ks, yielding roughly 10L counts within the $\mathrm{R}_{500}$ radius in 0.7-8.0 keV energy band. This depth helps us probe the thermodynamic structure on higher spatial resolution (scale of $\sim2\arcsec$). All \textit{Chandra} data were reprocessed using \textit{CIAO} v4.11 and \textit{CALDB} v4.8.2. The data reduction process and steps have been aggregated in a pipeline described in \citep{Datta2014,Schenck2014,Hallman2018,Alden_2019ascl.soft05022A}. It is designed to take \textit{Chandra} observation id's and generates temperature maps using three different temperature map-making techniques a) Adaptive Circular Binning (ACB), b) Weighted Voronoi Tessellation (WVT) Method, and c) Contour Binning Method (Contbin). The data were reprocessed using \textit{chandra\_repro} and cleaned the ACIS background in ``very faint'' mode.
As the cluster's extended emission ($\mathrm{R}_{500}\sim 8.49\arcmin$) fills all the ACIS-I chips, we decided to model the background contribution present in the observation from the ``blank-sky'' background files provided in the calibration database (CALDB). These  ``blank-sky'' background files represent the particle background as well as unresolved cosmic X-ray background.
These background files were reprocessed in the same way as the cluster data for each observation. We re-projected the events from the ``blank-sky'' files onto the same sky representing the cluster data to extract the background spectra from the same region used in the spectral analysis.
We remove intervals with background flares using light curves (259s bin) in the full energy band and the 9-12 keV band. We use \textit{deflare} tool to remove count rates greater than $3\sigma$ from the mean.  We inspect the light curve visually to ensure if flares were effectively removed.
All the point sources present in the cluster observations were identified using an automated tool \textit{wavdetect} inbuild in \textit{CIAO}. Point sources were identified in four different energy bands (0.5-1.2 keV, 1.2-2.0 keV, 2.0-7.0 keV, and 0.7-8.0 keV) with the scale of 1, 2, 4, 8, and 16pixels for careful subtraction. All the identified point sources were visually inspected for false detection or if \textit{wavdetect} failed to detect any real point sources before masking.
Regions with point sources were also masked from the background files to avoid negative subtraction. 

\begin{figure*}
\centering
\begin{tabular}{lccr}
\includegraphics[width=\columnwidth]{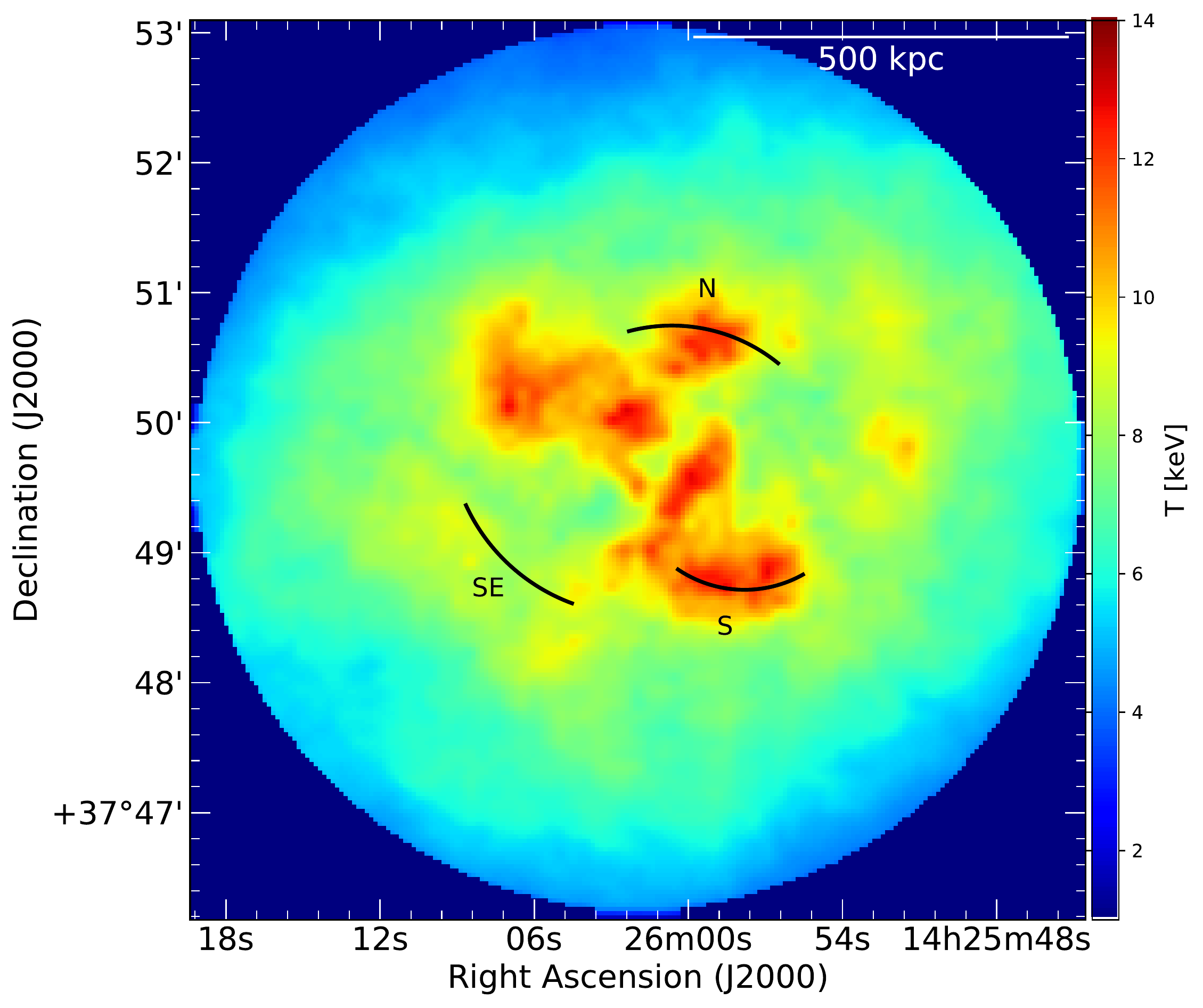} &
\includegraphics[width=\columnwidth]{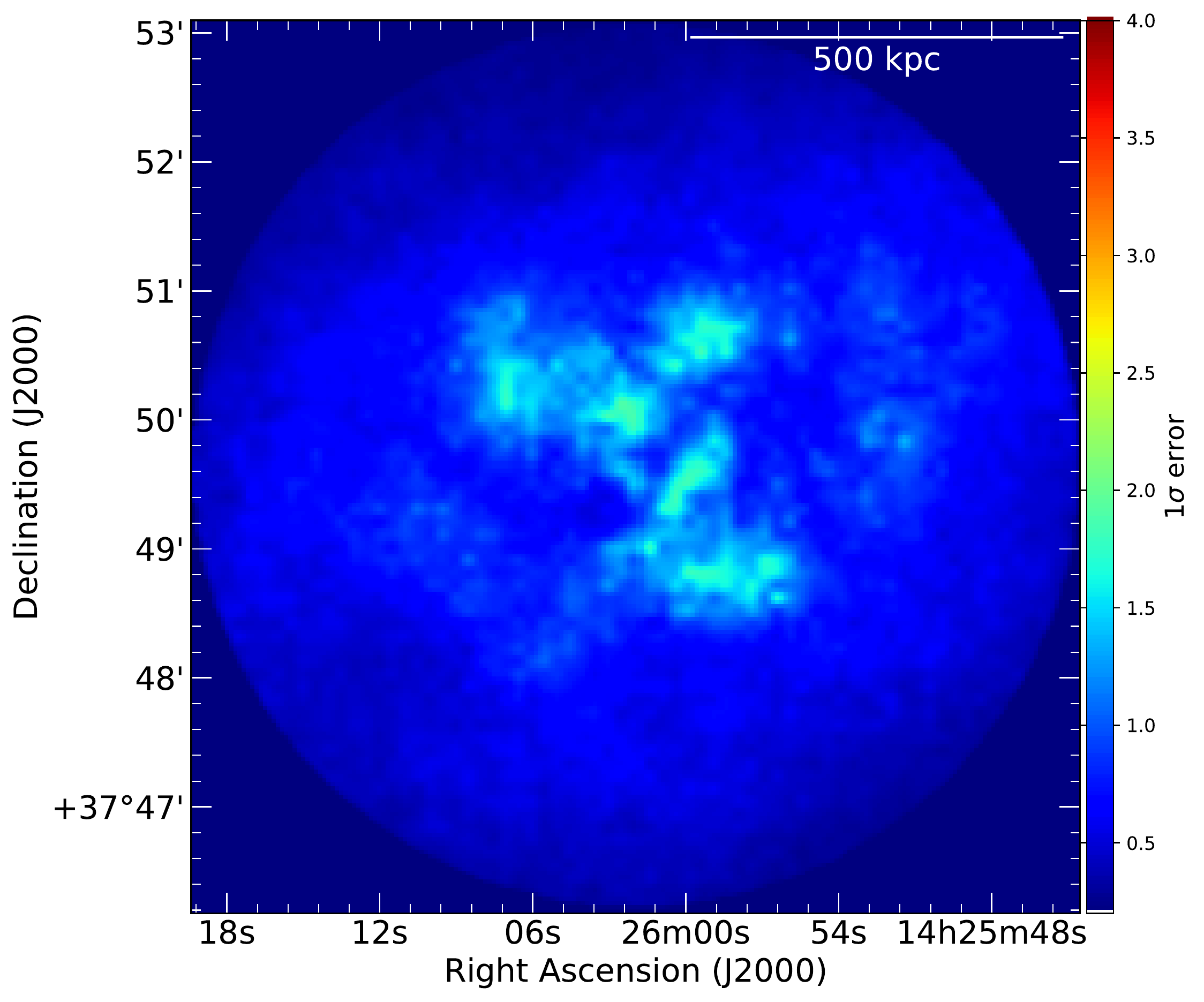} \\

\end{tabular}
\caption{\textit{Chandra} X-ray ACB temperature maps of the A1914. 
Due to the low counts within the outer region of the cluster, we were only allowed to make a temperature map of the inner 600kpc regions. Fractional error maps within 1$\sigma$ error are presented at the right panel.
To further investigate the significance of these temperature fluctuation, we have compared with temperature map of different techniques (see section \ref{WVT} and \ref{Contbin}) and plotted temperature profile (\ref{fig:NTjump_50_105}, \ref{fig:STjump_235_300}, and Figure \ref{fig:SETjump_204_250}) along the high temperature regions (wedge are shown in top left panel of Figure \ref{fig:X-Ray-SB}). Arcs (N, S, and SE in black) represent the position of the shock waves.
These structures also resemble the RGB image of the X-ray surface brightness map presented in the bottom left, Figure \ref{fig:X-Ray-SB}.}
\label{fig:ACB_Tmap}
\end{figure*}

\begin{figure*}
\centering
\begin{tabular}{lccr}
\includegraphics[width=\columnwidth]{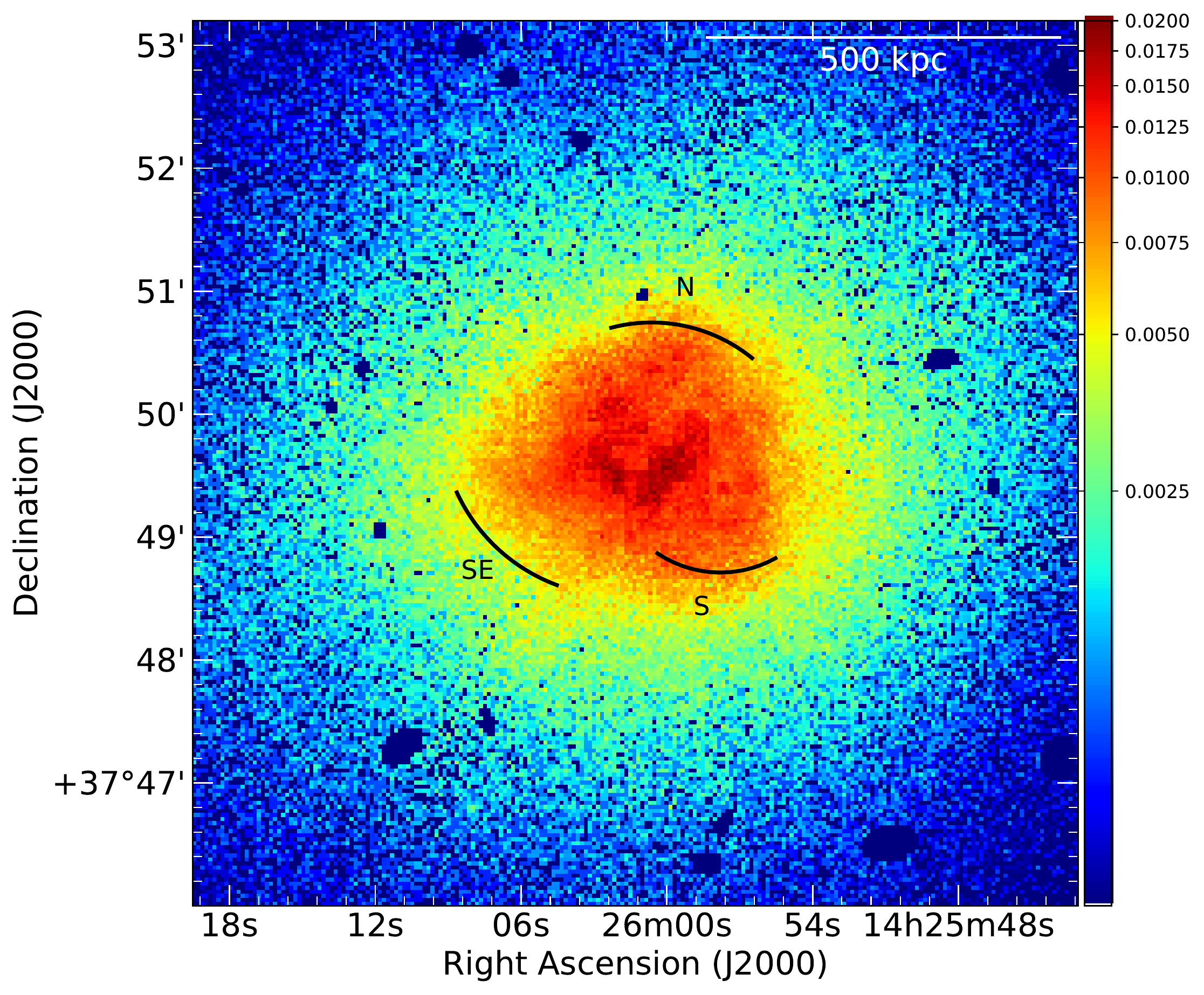} &
\includegraphics[width=\columnwidth]{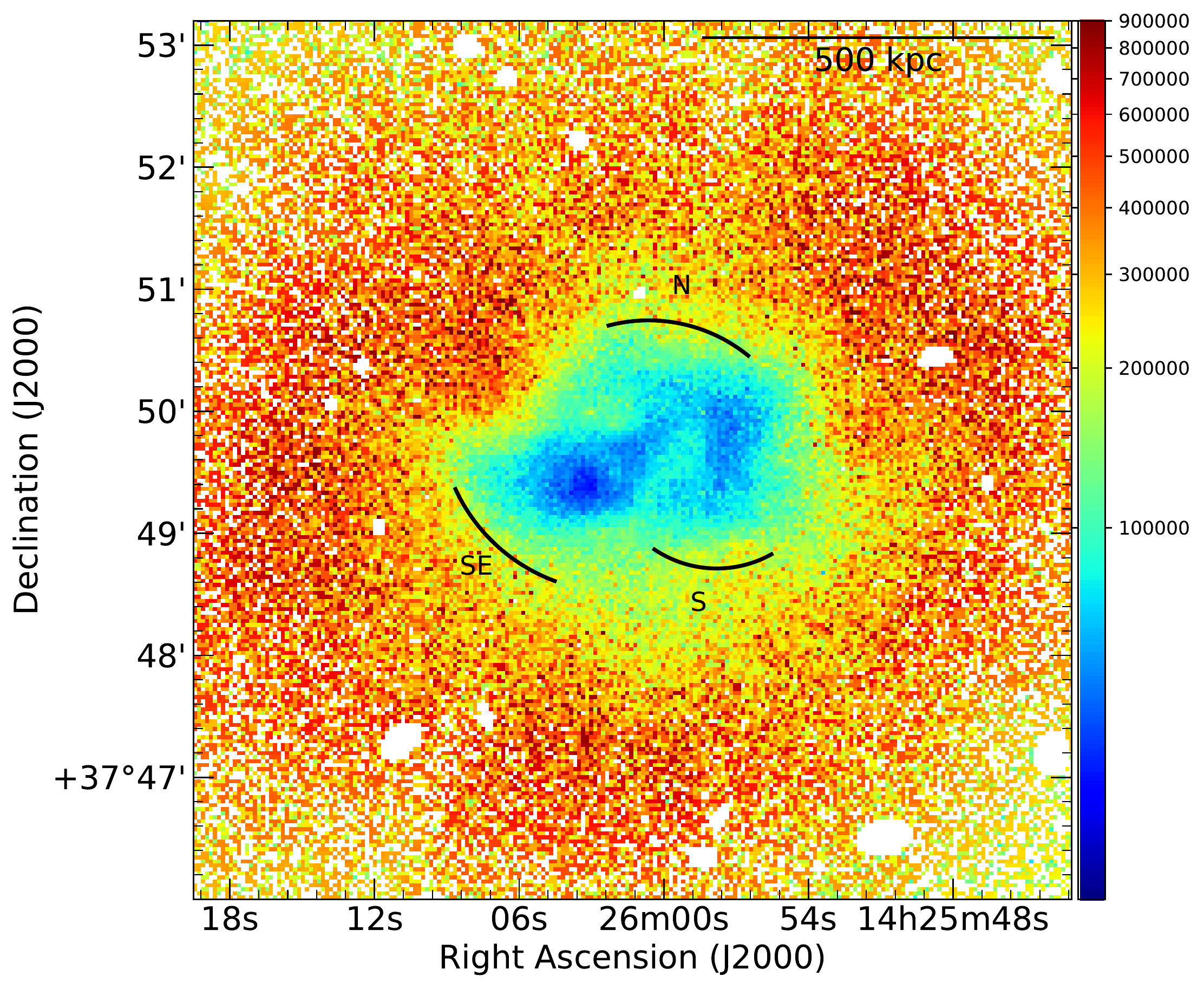} \\
\end{tabular}
\caption{Left: Pseudo pressure map of A1914 using $P = n_e kT \propto S_X^{1/2}kT$. Right: Pseudo entropy map calculated using $K = n_e^{-2/3}kT \propto S_X^{-1/3}kT$. Here, $n_e \propto S_X^{1/2}$, $S_X$ is the surface brightness shown in top left panel of Figure \ref{fig:X-Ray-SB} and $kT$ represents by ACB temperature map (top left panel of Figure \ref{fig:ACB_Tmap}).}
\label{fig:press_ent}
\end{figure*}

\begin{figure*}
\centering
\includegraphics[height=6.5in]{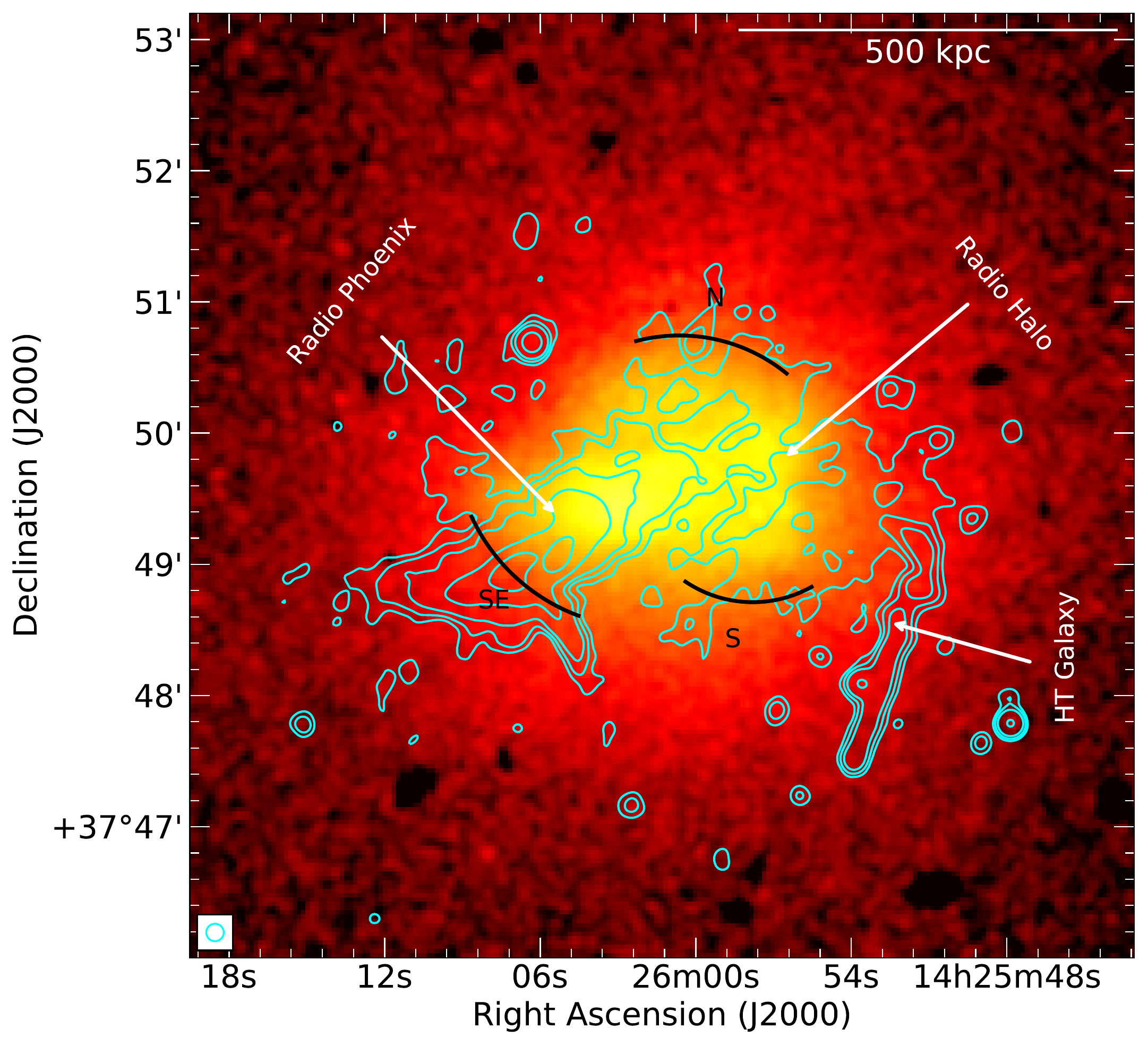}
\caption{\textit{Chandra} X-ray surface brightness map overlaid with the contour of 610 MHz GMRT radio image (cyan). The radio contours are drawn at levels $[-1,1,2,4,8,...]\times 3\sigma_{\mathrm{rms}}$, where $\sigma_{\mathrm{rms}}$= 65 $\mu$Jy beam$^{-1}$. The beam size of 610 MHz images is $8\arcsec \times 8\arcsec$  and is shown in the bottom left corner. We labeled radio phoenix, radio halo, and HT galaxy following the radio study by \citet{Mandal_a2019A&A...622A..22M}. Newly detected shock positions are also shown at N, S, and SE.}
\label{fig:Xsb_R600}
\end{figure*}

\begin{table}
\renewcommand{\arraystretch}{1.5}
\caption{Parameters of A1914.}
\label{tab:parameters}
\begin{tabular}{lcc}
\hline
Parameter & Value & Reference \\
\hline
RA (J2000) &  14:26:02.00 & 1 \\
Dec (J2000) &  +37:49:38.0 & 1 \\
Redshift (z) & $0.1678\pm 0.0004$ & 2  \\
$R_{500}$ & 1.52 (Mpc) & 3 \\
$M_{500}$ [$10^{14} \mathrm{M_{\odot}} $] & $11.5 \pm 1.8$ & 3 \\
\hline
\end{tabular}
\\References: 1. \citet{Botteon_2018MNRAS.476.5591B}, 2. \citet{Barrena2013MNRAS.430.3453B} 3. \citet{Maughan2016MNRAS.461.4182M} 
\end{table}

\subsection{Spectral Analysis} \label{sec:spectral_analysis}
X-ray spectroscopy is an essential tool to obtain insight into the structure and astrophysics of galaxy clusters.
Imaging and spectroscopy of X-ray extended sources require a proper characterization of a spatially unresolved background signal.
As A1914 is a low redshift cluster, its emission fills all the ACIS-I chips.
Therefore, to get a more accurate model of the non-sky quiescent background, we used ``bank-sky'' background files available in the \textit{Chandra} calibration database.
This ``bank-sky'' background were re-normalized in the 9.5-12 keV band; because, at these energies, the \textit{Chandra} effective area is negligible, and all the 9.5-12 keV flux in the sky data is due to particle background \citep{Hickox2006ApJ...645...95H}.

In order to create the best quality temperature maps of a cluster of galaxies, we applied three different temperature map-making techniques: (1) Adaptive Circular Binning (ACB), (2) Weighted Voronoi Tessellation Method (WVT, see section \ref{WVT}), and (3) Contour Binning Method (Contbin, see section \ref{Contbin}) method. In each method, temperature maps were created using APEC and PHABS  thermal plasma model. APEC is an emission spectrum from collisionally-ionized diffuse hot gas, and PHABS is a photoelectric absorption model. We have used redshift z = 0.168, metallicity (Z) = 0.3 $Z_\odot$, and column density $ (N_H) = 9.87 \times 10^{19}cm^{-2}$ \citep{HI4PI_2016A&A...594A.116H}. 
We used the threshold signal-to-noise (SNR) criterion to have a minimum count per region or bin for each method. Here, the signal is CLEAN data, and noise assumes Poisson's uncertainty contribution from both the CLEAN and blank-sky backgrounds. Since the SNR drops at the periphery of the cluster, we only made a temperature map of the 600 kpc inner region for all three different types of temperature maps.

Various temperature map-making techniques have their own advantages and limitations. The WVT and Contbin methods create non-overlapping regions to extract the spectrum. The best fitted temperature values and the associated errors are independent of those in the neighboring regions. As the Contbin method reliably creates bins that follow the surface brightness, it is preferable to search for shocks or cold fronts within the cluster region \citep{Sanders2006}. The major limitation of WVT and Contbin methods is that the resultant temperature map lacks resolution and is heavily depends on the bin locations. These bin locations are not unique and change due to the initial conditions of the bin accretion. A slight change in the input SNR value or the starting location of the bin-accretion can create slightly different temperature maps. In contrast, an ACB region significantly overlaps with the neighboring ACB regions. The area of the overlap regions depends on the SNR around the reference pixel. 

\subsubsection{Adaptive Circular Binning (ACB) Method} \label{sec:ACB}
ACB temperature map was produced by using a method described in \citet{Datta2014,Schenck2014,Hallman2018,Alden_2019ascl.soft05022A,Rahaman_2021MNRAS.505..480R}. Spectra were extracted from the circular region centered on every pixel.
We used the SNR = 60 to generate circular ACB regions. In this process, circles are allowed to overlap with each other.
\textit{CIAO} task \textit{dmextract} was used to create spectra from each region from each observations and \textit{specextract} was used to calculate response matrices (RMF and ARF). 

We have performed spectral fitting using XSPEC version: 12.9.1 in between 0.7-8.0 keV energy range. The APEC and PHABS models were fitted to the spectra from each region simultaneously using C-statistics \citep{Cash1979}. The metallicity of the cluster was kept frozen at 0.3 $Z_\odot$ throughout the cluster, where $Z_\odot$ is the solar abundance in \citep{Anders_1989GeCoA..53..197A}. Redshift (0.168; \citealt{Barrena2013MNRAS.430.3453B}) and $N_H$ ($9.87 \times 10^{19}cm^{-2}$; \citealt{HI4PI_2016A&A...594A.116H}) were also kept frozen. Only APEC normalization and temperature were fitted for each region. The best-fitted temperature and errors of each circular spectral region were assigned to the center of the circle. 
The temperature and corresponding error maps are obtained (Figure \ref{fig:ACB_Tmap}).

\subsection{X-ray Imaging Analysis}
\subsubsection{Unsharp masking}
An ideal relaxed cluster shows spherical symmetry; however, a non-relaxed (merging) galaxy cluster deviates from that.
Such deviations can be seen in the residual of the surface brightness (SB) map of any disturbed clusters. 
The 0.7-8.0 keV \textit{Chandra} X-ray surface brightness image of A1914 was used to produce an unsharp masked image, an image sharpening technique often used in digital image processing to unveil underlying structures. At first, we smoothed the image with a 2$\sigma$ wide Gaussian kernel using \textit{aconvolve}, a task within \textit{CIAO}. This helps to suppress the pixel-to-pixel fluctuations while preserving small-scale uncorrelated structures of the X-ray brightness. Secondly, we also smoothed the image with a 10$\sigma$ wider Gaussian kernel, which erases small-scale features while conserving the overall morphology of the hot gas distribution within the A1914. Finally, we subtract the image smoothed with 10$\sigma$ from that smoothed with 2$\sigma$  wide Gaussian kernel to generate the unsharp masked image. The top right panel of Figure \ref{fig:X-Ray-SB} shows the resulting unsharp masked image.
A thorough inspection of this figure strengthens the presence of X-ray surface brightness edges in the ICM of A1914. 

\subsubsection{Edge detection filter (GGM)}
The visual search of X-ray surface brightness images allows
identifying the apparent discontinuities \citep{Markevitch2007}. 
We accompany this approach with the visual inspection of the Gaussian gradient magnitude (GGM) filter image.
\citet{Sanders2016MNRAS.460.1898S} presented a GGM filter that aims to highlight the SB gradients in an image. The GGM filter calculates the gradient of an image assuming Gaussian derivatives with width $\sigma$ \citep{Sanders2016MNRAS.460.1898S}.
These GGM images help identify sharp candidate edges, such as shocks and cold fronts \citep{Walker2016MNRAS.461..684W,Botteon_2018MNRAS.476.5591B}.
We applied this GGM filter to the \textit{Chandra} X-ray surface brightness map, and the output image is shown in the bottom right panel of Figure \ref{fig:X-Ray-SB}, where we used a GGM filter with a width of $\sigma$ = 4 pixels (where a pixel corresponds to 0.984 $\arcsec$). 

\subsubsection{RGB Chandra X-ray image}
We made a RGB tricolor image from \textit(Chandra) X-ray surface brightness map within the range of 0.7-8.0 keV.  We generate X-ray surface brightness maps in three different energy bands, namely, soft (0.7-1.2 keV), medium (1.2-2.0 keV), and hard (2-8.0 keV). All these images were smoothed with a Gaussian kernel of $\sigma$ = 3 pixels and combined to generate an RGB image (bottom left panel of Figure \ref{fig:X-Ray-SB}). The red color in this image represents the hard component of the X-ray emitting gas, while the medium and soft components are represented in green and blue, respectively. This tricolor image represents the signature of the broad spectral properties of the X-ray emitting gas. Here, this RGB image highlights the presence of several substructures in A1914.

\section{Radio data analysis}
The 610 MHz GMRT observation of A1914 used here (31\_049), was done on 4/12/2016 for about 5.7 hrs. The observation was carried out in dual-polarization mode with 32 MHz bandwidth divided into 256 spectral channels.

The data reduction was performed using SPAM, which does RFI (Radio Frequency Interference) mitigation, image plane ripple suppression, and direction-dependent calibration to deal with ionospheric corruptions \citep{Intema2009A&A...501.1185I,Intema2017A&A...598A..78I}. The image shown in Figure \ref{fig:Xsb_R600} was created using Briggs robust parameter = 0 \citep{Briggs1995}, and was later smoothed with a Gaussian beam of $8\arcsec\times8\arcsec$.

Here, we presented this radio map for spatial comparison between X-ray and radio morphology. As this article focuses on X-ray study, we refer readers to \citet{Mandal_a2019A&A...622A..22M} for a detailed radio study on this cluster.

\section{Results} \label{results}
\subsection{Thermodynamic maps}
To investigate the thermodynamic properties of the cluster, we created thermodynamic maps (e.g., SB, temperature, pressure, and entropy maps) of the cluster.
The top left panel of Figure \ref{fig:X-Ray-SB} shows the merged, flat-fielded (vignetting and exposure corrected), and background-subtracted 0.7-8.0 keV band Chandra ACIS-I image of A1914. The distribution of X-ray emitting gas has a complex morphology. 
The gas distribution is asymmetric and elongated in the Southeast and Northwest direction. This image and other residual images in Figure \ref{fig:X-Ray-SB} show hints of surface brightness edges in the cluster. 
To explore the nature of these features and constrain the merger history, we derive temperature and density profile (see section \ref{jump_model}) as well as thermodynamic maps of the cluster.
The temperature map of the cluster using three different techniques are shown in Figure \ref{fig:ACB_Tmap} and \ref{fig:wvt_contbin_Tmap}.
All three maps are consistent with each other and have the same disturbed temperature structure.
We have used more archival Chandra observations, which resulted in a significantly improved temperature map, covering a larger cluster region compared to the previous studies  \citep{Govoni_2004ApJ...605..695G,Botteon_2018MNRAS.476.5591B}.

For clusters with typical temperatures (kT = 3-10 keV), the broadband response of \textit{Chandra} assumed to be constant (about 17\% less at 4-12 keV with respect to count rate at 0.5-2.5 keV, \citealt{Santos2019ApJ...887...31A}).
If one ignores the \textit{Chandra} response, the gas density is proportional to the square-root of the count rate per unit volume of the gas.
Thus, we can map the projected density of the cluster from the surface brightness.
Furthermore, combining the projected density with a temperature map, we can compute the pseudo pressure and pseudo entropy maps using the following relations:

\begin{ceqn}
\begin{align} \label{eq:ne}
    n_e \propto S_X^{1/2},
\end{align}
\begin{align} \label{eq:P}
    P = n_e kT \propto S_X^{1/2}kT,
\end{align}
\begin{align} \label{eq:K}
    K = n_e^{-2/3}kT \propto S_X^{-1/3}kT,
\end{align}
\end{ceqn}

Where $S_X$, T, P, and $K$ are the surface brightness, projected temperature, projected pressure, and projected entropy, respectively. 
Figure \ref{fig:press_ent} shows the pseudo-pressure and pseudo-entropy maps which confirms the disturbed morphology of the cluster.

\subsection{Searching for Shocks}
We explored different analysis approaches of X-ray observations to detect and characterize edges of massive cluster A1914. We used both imaging and spectral analysis to investigate the shock detected in the cluster.
From all the images represent in Figure \ref{fig:X-Ray-SB} and \ref{fig:ACB_Tmap}, one can pinpoint the complex morphology of the ICM, SB, and temperature variations within the central region of the cluster. Spatial variation in the central region can also be seen in pressure and entropy maps in Figure \ref{fig:press_ent}.
These can be thus investigated with the fitting of SB profiles and temperature profiles (see section \ref{sec:therm_prof}), whose extracting sectors have to be accurately chosen to describe the putative shock or cold front properly. 
Once the edge is well located, one can efficiently perform spectral analysis on dedicated upstream and downstream regions. 
To investigate the origin of these SB edges, we search for SB discontinuity in the \textit{Chandra} X-ray emission by extracting SB profiles (within 0.5-2.0 keV energy band) using PROFFIT v1.5 \citep{Eckert2011A&A...529A.133E,Eckert2016ascl.soft08011E}. 
We used 2$\arcsec$ (arcsec) binning to extract the surface brightness profile to have enough count in each bin.
We find discontinuities in density profile over the wedge regions N, S, and SE (shown in top left of Figure \ref{fig:X-Ray-SB}).
We also check for the discontinuities in the temperature profile over the same region using the classical method described in section \ref{sec:therm_prof}.

\begin{figure*}
    \centering
    \includegraphics[height=\columnwidth]{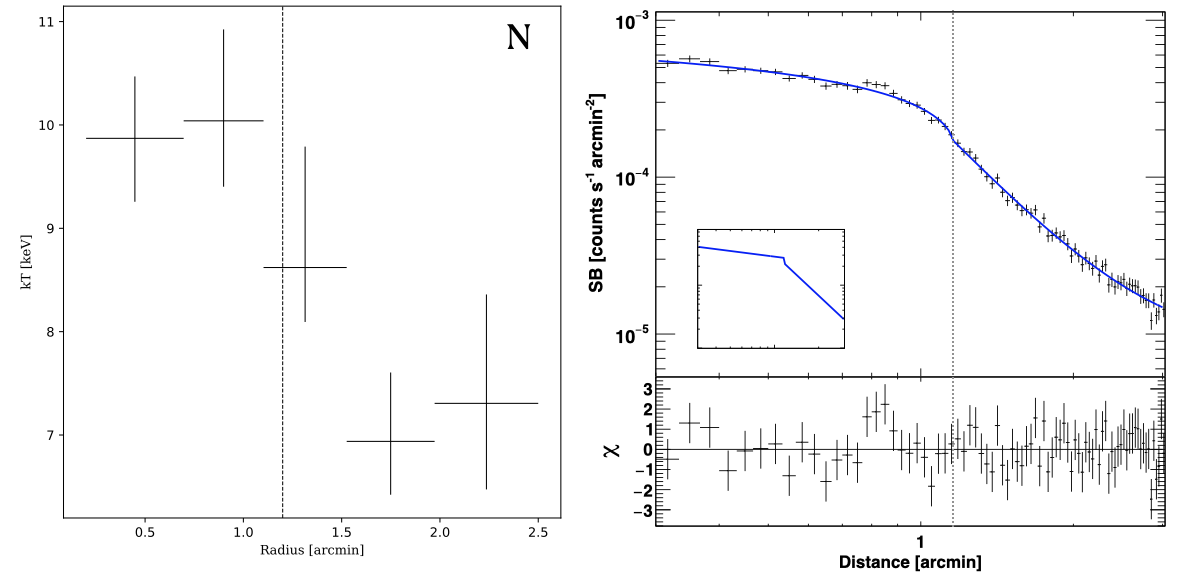} \\
    \includegraphics[width=6.9in]{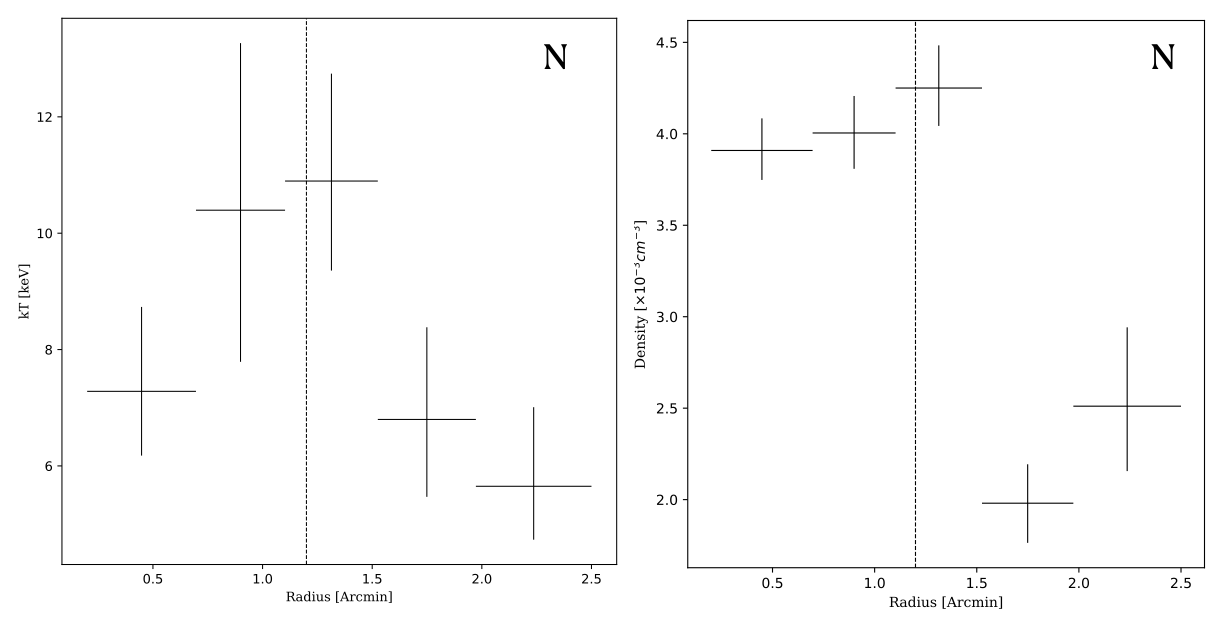}
    \caption{Top: Projected \textit{Chandra} X-ray temperature (left) and Surface brightness (right) profiles across the N wedge region (N, opening angle 50-105 deg) shown in Figure \ref{fig:Xsb_R600}. We used 2$\arcsec$ binning for surface brightness profile. The broken power-law model (blue line) was fitted to surface brightness profile using Proffit v1.5 (see section \ref{jump_model}). The inset in the top right panel shows the simulated gas density model. The bottom of the top right panel shows the residuals obtained from the fit. Best fitted parameters are listed in Table \ref{tab:djump}.
    Bottom: Deprojected \textit{Chandra} X-ray temperature (left) and density (right) profile over the same region.
    The distances (in arc-min) are not from the center of the cluster but the center of the wedge region.}
    \label{fig:NTjump_50_105}
\end{figure*}

\begin{figure*}
    \centering
    \includegraphics[height=\columnwidth]{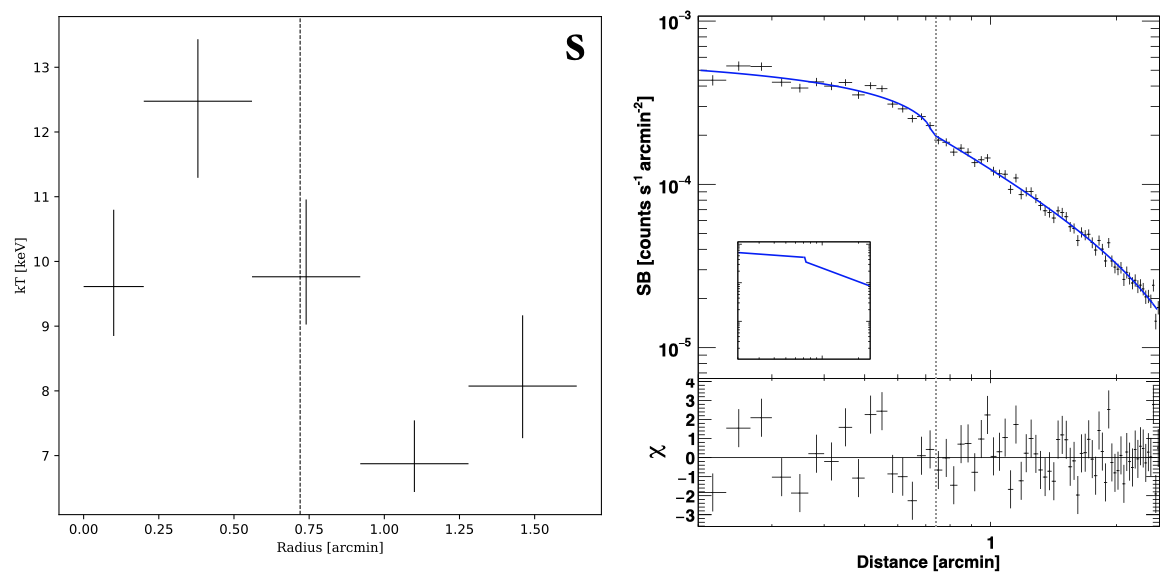} \\
    \includegraphics[width=6.9in]{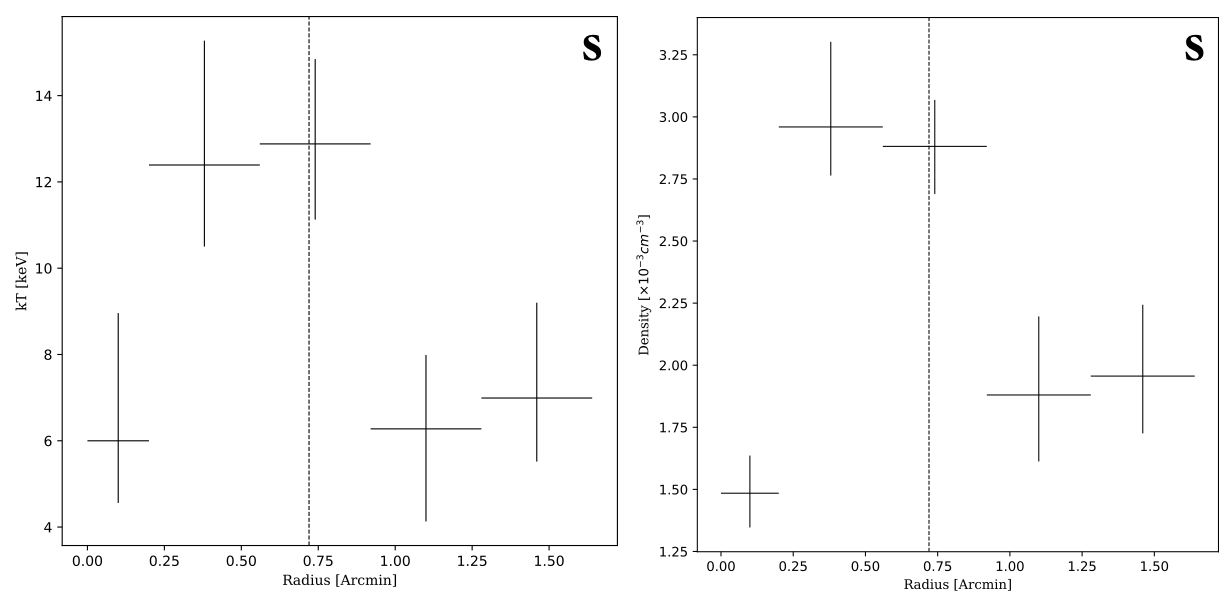}
    \caption{ Top: Projected \textit{Chandra} X-ray temperature (left) and Surface brightness (right) profiles across the S wedge region (S, opening angle 235-300 deg) shown in Figure \ref{fig:Xsb_R600}. We used 2$\arcsec$ binning for surface brightness profile. The broken power-law model (blue line) was fitted to surface brightness profile using Proffit v1.5 (see section \ref{jump_model}). The inset in the top right panel shows the simulated gas density model. The bottom of the top right panel shows the residuals obtained from the fit. Best fitted parameters are listed in Table \ref{tab:djump}.
    Bottom: Derojected \textit{Chandra} X-ray temperature (left) and density (right) profiles over the same region.}
    \label{fig:STjump_235_300}.
\end{figure*}

\begin{figure*}
    \centering
    \includegraphics[height=\columnwidth]{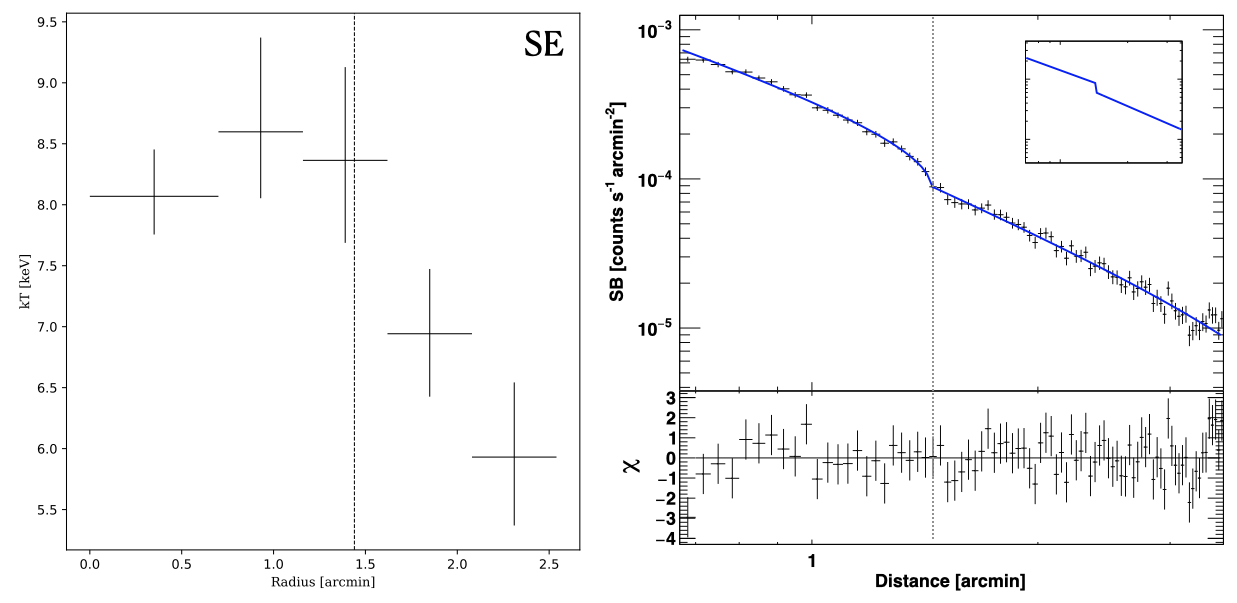} \\
    \includegraphics[width=6.7in]{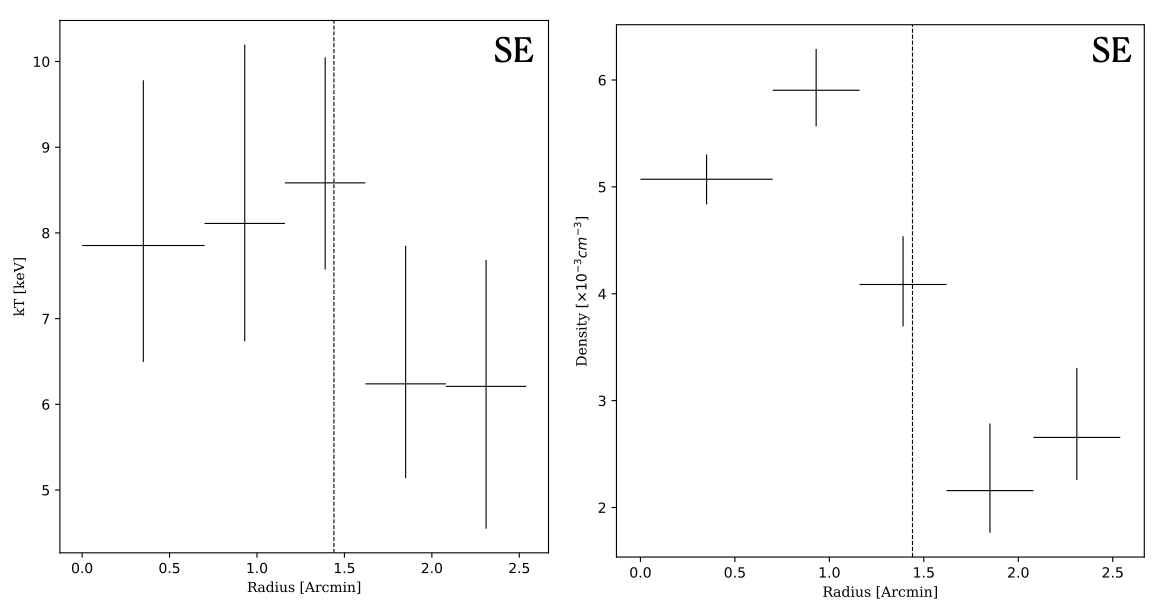}
    \caption{Top: Projected \textit{Chandra} X-ray temperature (left) and Surface brightness (right) profiles across the SE wedge region shown in Figure \ref{fig:Xsb_R600}. We extracted these profiles over the wedge region SE with an opening angle of 204-250 deg.
    We used 2$\arcsec$ binning for surface brightness profile. The broken power-law model (blue line) was fitted to surface brightness profile using Proffit v1.5 (see section \ref{jump_model}). The inset in the top right panel shows the simulated gas density model. The bottom of the top right panel shows the residuals obtained from the fit. Best fitted parameters are listed in Table \ref{tab:djump}.
    Bottom: Derojected \textit{Chandra} X-ray temperature (left) and density (right) profiles over the same region.
    }
    \label{fig:SETjump_204_250}
\end{figure*}

\begin{figure*}
\centering
\includegraphics[width=7in]{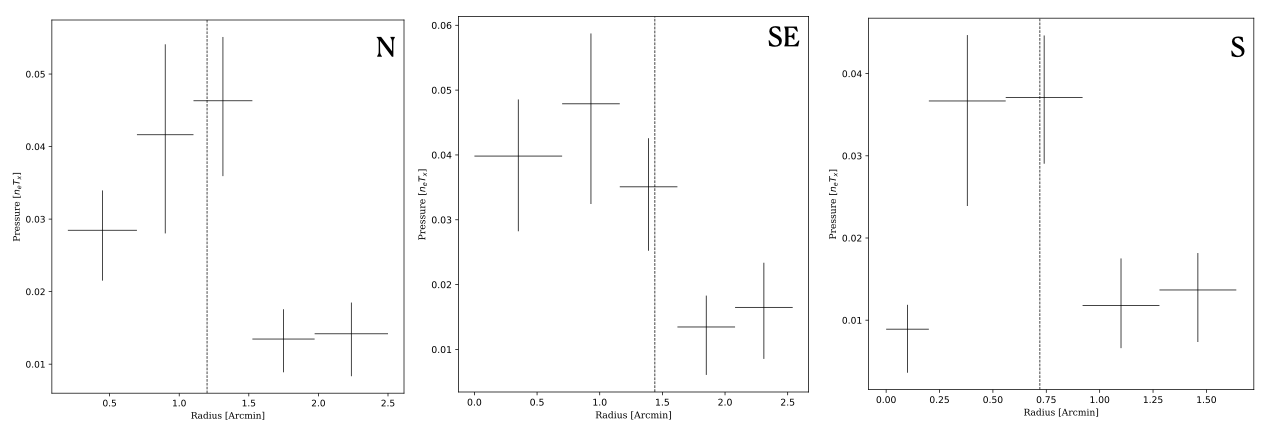}
\caption{These are the deprojected pressure profiles over the three wedge region labeled as N, SE, and S in Figure \ref{fig:X-Ray-SB}. We used $P = n_e kT$ to calculate the pressure profile, where $kT$ is deprojected temperature and $n_e$ is the deprojected density.}
\label{fig:all_pressure}
\end{figure*}

\begin{table*}
    \centering
    \begin{tabular}{ccccccccc}
        \hline
        Sector & $\alpha1$ & $\alpha2$& $\mathrm{r_{shock}}$ (arcmin) & Jump (C) & $red-\chi^{^2}$ & $M_{SB}$ & $red-\chi^{^2}$ & $red-\chi^{^2}$ \\
        & & & & & Broken power law & & Power law & Beta Model \\
        \hline
        N & $0.29\pm 0.036$ & $2.18\pm 0.073$ & $1.15 \pm 0.010$ & $1.28\pm 0.05$ & 70.14/75 & $1.19^{+0.03}_{-0.03}$ & 1259.65/77 & 193.49/77  \\
        S  & $0.22\pm 0.064$ & $1.16 \pm 0.044$& $0.72 \pm 0.009$ & $1.29 \pm 0.049 $ & 95/63 & $1.18^{+0.035}_{-0.032}$ & 355/65 & 144/65  \\
        SE & $1.3\pm 0.053$ & $1.6 \pm 0.091$ & $1.44 \pm 0.007$  & $1.42 \pm 0.052$ & 66.3/78 & $1.28^{+0.04}_{-0.02}$ & 225.5/80 & 155/80 \\
        \hline
    \end{tabular}
    \caption{Best fitted parameters of the broken power-law model (bknpow, fitted using PROFFIT v1.5) are listed here. Columns 2-5 are the best-fitted parameters of the broken power-law model. The Mach numbers from corresponding density jumps are also calculated in column 7 using equation \ref{eq:den2}. Column  8 and 9 represents the reduced-$\chi^{^2}$ of single power-law and beta model respectively.}
    \label{tab:djump}
\end{table*}

\begin{table}
    \centering
    \begin{tabular}{cccc}
        \hline
        Sector &  $T_{down}$ (keV) &  $T_{up} (keV) $ & $M_T$ \\
        \hline
        N & $10.9^{+1.85}_{-1.53}$ & $6.8^{+1.58}_{-1.33}$ & $ 1.6^{+0.56}_{-0.48}$ \\
        S & $12.9^{+1.96}_{-1.75}$ & $6.3^{+1.71}_{-2.14}$ & $1.98^{+0.96}_{-0.58}$ \\
        SE & $8.6^{+1.46}_{-1.01}$ & $6.2^{+1.61}_{-1.10}$ & $1.4^{+0.52}_{-0.40}$ \\
        \hline
    \end{tabular}
    \caption{Deprojected temperature of the both side (Up and Downstream) of the shock are listed here. The Mach numbers are calculated using equation \ref{eq:RH}.}
    \label{tab:Tjump_t}
\end{table}

\subsubsection{Modeling the density jumps} \label{jump_model}
The extracted SB profiles (N, S, and SE) were fitted with a broken power-law density model. In all cases, discontinuities are evident within the 90\% confidence level. 
The broken power-law density model is defined as (Eq. \ref{eq:bknpow}): 

\begin{equation} \label{eq:bknpow}
  n(r) = \begin{cases}
    {\cal C} n_0(\frac{r}{r_{sh}})^{-\alpha 1} & \text{if $r < r_{sh}$} \\
    n_0(\frac{r}{r_{sh}})^{-\alpha 2}          &  \text{if $r > r_{sh}$}
  \end{cases}
\end{equation}

Where n is the electron number density, $n_0$ is the density normalization constant, {\cal C} is the compression factor, r is the radial distance from the center, $r_{sh}$ is the radial distance of the shock/cold fronts, and $\alpha 1\ \&\ \alpha 2$ are the power-law indices.

All the parameters of a broken power-law model were allowed to vary during the fit, and the best-fitted parameters are listed in Table \ref{tab:djump}. In order to compare the significance of the  broken power-law fit to the surface brightness profile, we have introduced fitting with single power-law and a beta model. We tabulated the final reduced-$\chi^2$ values for all the three models in Table \ref{tab:djump}. These results clearly shows that the surface brightness profile near all the three proposed shock locations is only consistent with the broken power-law model. This confirms further the existence of the density jump across all three proposed shock front locations.  However, to confirm whether the density jump is due to shock front or a cold front, we investigate further the nature of the temperature jump in these proposed locations (section \ref{sec:therm_prof}).

We calculate the Mach number (if discontinuity is a shock) from compression ratio {\cal C} via the Rankine-Hugoniot equations presented in Eq. \ref{eq:den1} 
\begin{ceqn}
\begin{align} \label{eq:den1}
    \frac{\rho_2}{\rho_1} = {\cal C} = \frac{(1 + \gamma)M^2}{2 + (\gamma - 1) M^2}
\end{align}
\end{ceqn}

For mono-atomic gas, $\gamma = 5/3$, Eq. \ref{eq:den1} becomes,

\begin{ceqn}
\begin{align} \label{eq:den2}
    {\cal C} = \frac{4M^2}{3 + M^2}
\end{align}
\end{ceqn}
The calculated Mach numbers are listed in Table \ref{tab:djump}.

\subsubsection{Non-detection of west shock}

\begin{figure*}
\centering
\includegraphics[width=7in]{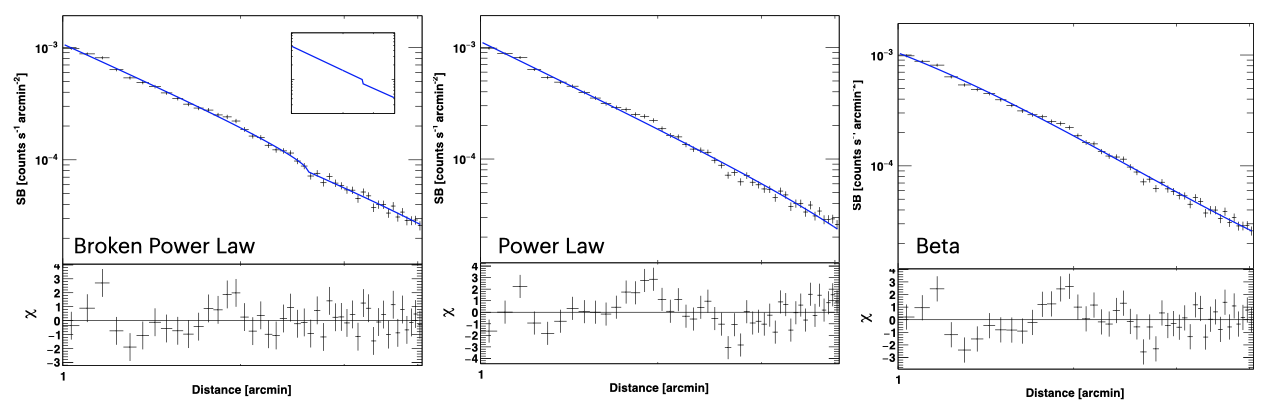}
\caption{\textit{Chandra} X-ray surface brightness profile across the wedge region towards the west of the center (between 330-380 deg, centered at 216.5094869, 37.8226528) where previously a shock was detected by \citep{Botteon_2018MNRAS.476.5591B}. Here, we show the surface brightness profile fitted with a broken power-law (left), single power-law (middle), and a beta model (right). The best-fitted parameters of each model are listed in Table \ref{tab:w-param}.}
\label{fig:W-wedge}
\end{figure*}

We extensively search for the discontinuity at the shock position reported by \citet{Botteon_2018MNRAS.476.5591B}. We could not find any discontinuity in surface brightness at the proposed location towards the west region (between 330-380 deg, centered at 216.5094869, 37.8226528). We investigated the discontinuity by fitting the profile with a broken power law, a single power law, and a beta model. All the models fitted well in the surface brightness profile (see Figure \ref{fig:W-wedge}), which implies no discontinuity over this region. The best-fitted parameters for all the models are listed in Table \ref{tab:w-param}.

\begin{table}
    \centering
    \begin{tabular}{ccc}
        \hline
        \hline
         Model & Parameters &  $\chi^{^2}$/d.o.f. \\
         \hline
         Broken power law & $\alpha_1 = 1.67 \pm 0.034 $ & 43.7/39   \\
         & $\alpha_2 = 1.58 \pm 0.432 $ & \\
         & $r_{shock} = 2.61 \pm 0.073$ arcmin & \\
         & Jump (C) = $1.2 \pm 0.049$ & \\
         \hline
         Single power law & $\alpha = 2.54 \pm 0.041 $ & 76.1/41 \\
         & $rs = 1.01 \pm 0.034 $ & \\
         & $norm = 1.0 \pm 0.092\times 10^{-3}$ & \\
         \hline
         Beta & beta = $0.72 \pm 0.05$ & 64.29/41 \\
         & rc = $0.80 \pm 0.15$ & \\
         & norm = $5.05 \pm 1.39 \times 10^{-3}$ & \\
         \hline
    \end{tabular}
    \caption{The best-fitted parameters of the broken power law, single power law, and beta model fitted in the surface brightness profile over the west wedge region are listed here.}
    \label{tab:w-param}
\end{table}

\subsubsection{Deprojected thermodynamic profiles across the SB edges } \label{sec:therm_prof}

Observed thermodynamic profiles in a galaxy cluster are affected by projection effect. The variation in the thermodynamic quantities may get smoothed out due to these projection effects. Therefore, we performed deprojection analysis over all the wedge regions (shown in Figure~\ref{fig:X-Ray-SB} top-left). We assume spherical symmetry to deproject the temperature profile across the three shock front locations.  Both the projected and deprojected \textit{Chandra} X-ray temperature profiles over three wedge regions are shown in Figure \ref{fig:NTjump_50_105},  \ref{fig:STjump_235_300}, and \ref{fig:SETjump_204_250}. Each of the wedge regions were splitted into five annuli. We extract spectra from each annuli using \textit{dmextract} (\textit{CIAO} task) and  fitted with \textit{phabs*apec} for projected and \textit{projct(phabs*apec)} model for deprojected profile using XSPEC v:12.9.1  \citep{Arnaud1996ASPC..101...17A} in 0.7-8.0 keV energy range same as the case of temperature map. The temperature profiles, corresponding to these three shock locations, are shown with $\pm 1\sigma$ uncertainty. One should note that, due to artifacts introduced during the fitting of the projct model, the fitted temperatures in the outermost bins of the deprojected profiles are slightly lower than the projected values in the same bin.
All these temperature profiles show significant temperature discontinuity with gradient in the same direction as the surface brightness jumps.
To check the significance of these jumps, we also made pressure profiles for each wedge region (see Figure \ref{fig:all_pressure}) using deprojected temperature ($kT$) and density ($n_e$) profiles. We used $P = n_e kT$ to calculate pressure profiles.
These profiles further strengthen the probabilities of these edge to be shock fronts. The jump in the pressure profile of the SE edge is minor, but pointing to be a shock front.
Therefore, we consider SE-edge as a tentative detection of a minor shock perpendicular to the merging axis. We need deep X-ray observations to further investigate the SE edge.

We used the Rankine-Hugoniot temperature jump condition (Eq. \ref{eq:RH}) to calculate the Mach number.
\begin{equation}
\centering
    \frac{T_2}{T_1}=\frac{(5M^2-1)(M^2+3)}{16M^2}.
    \label{eq:RH}
\end{equation}
Where, $T_1$ and $T_2$ are the pre-shock and post-shock temperature, respectively. Temperature in \textit{up} and \textit{down}-stream along with calculated Mach numbers for all the sectors are listed in Table \ref{tab:Tjump_t}.
In addition, we calculated the deprojected density profile from the {\it apec} normalization constant ($N$) using equation \ref{eq:norm}.

\begin{equation} \label{eq:norm}
    N = \frac{10^{14}}{4\pi [D_A(1+z)]^2} \int n_e n_H dV \ \ \ (cm^{-5})
\end{equation}

where, $D_A$ is the angular diameter distance to the source (in cm), z is the redshift of the cluster, $dV$ is the volume of the extracted region, and $n_e$ and $n_H$ are the electron and hydrogen densities ($cm^{-3}$), respectively, and we assume $n_e$ = 1.18 $n_H$. Deprojected density profiles for all the three wedge sectors are shown in bottom right panel of Figure \ref{fig:NTjump_50_105},  \ref{fig:STjump_235_300}, and \ref{fig:SETjump_204_250}. This further strengthens the claim of all the three shock fronts.

\begin{figure}
    \centering
    \includegraphics[width=\columnwidth]{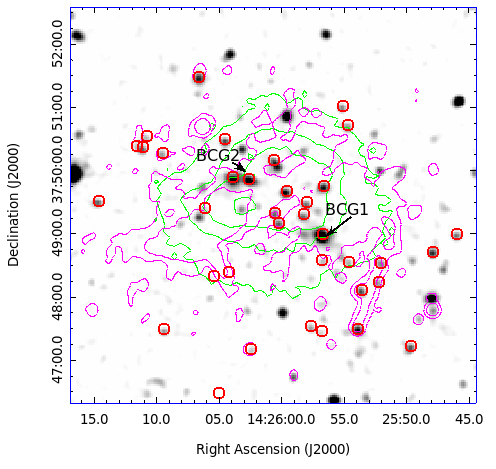} 
    \caption{SDSS r-band optical image overlaid with the contour of 610 MHz GMRT radio image (magenta). The radio contours are drawn at levels $[3,14,360]\times \sigma_{\mathrm{rms}}$, where $\sigma_{\mathrm{rms}}$= 65 $\mu$Jy beam$^{-1}$. \textit{Chandra} X-ray contour are shown in green. Red circles are galaxies taken from SDSS-DR12 catalog within the cluster redshift range. BCG1 and BCG2 are labeled as described in \citet{Barrena2013MNRAS.430.3453B}.}
    \label{fig:Optical}
\end{figure}

\section{Discussions}  \label{Discussions}
\subsection{Properties of the shock fronts}
We estimate the amplitude of the surface brightness jump from the broken power-law model (i.e., the surface brightness ratio in front and behind the shock) fitted to the surface brightness profile. 
We also estimate the jump in the temperature profile. From these jump information, the Mach numbers of the shocks are estimated using the Rankine Hugoniot jump conditions (see equation \ref{eq:den2}, \& \ref{eq:RH}). The results are shown in Table \ref{tab:djump} \& \ref{tab:Tjump_t}.
We include a 10\% systematic error during the spectral fitting in the temperature profile.
The calculated Mach numbers are listed in Table \ref{tab:djump} \& \ref{tab:Tjump_t}.
All three shocks are weak, with Mach number ranges from 1.13-1.64.
Mach numbers detected from density jumps are slightly lower (within 1$\sigma$ uncertainty) than the corresponding one seen using temperature discontinuity due to projection effect and systematic error.
At the place of SE shock, \citet{Botteon_2018MNRAS.476.5591B} reported a cold front using old shallow observation (21 ks) from \textit{Chandra}. There was no sign of temperature discontinuity in their study as the uncertainty was high due to low counts.
We re-investigate it with the recent deep \textit{Chandra} archival observations (146 ks) and found no sign of a cold front. 
However, we found discontinuities in both density and temperature profiles. We also create pressure profile from the deprojected temperature and density profile (see Figure \ref{fig:all_pressure}), which also shows discontinuity with gradient similar to the temperature and density jump.
As the gradient of the discontinuities is in the same direction (see Figure \ref{fig:SETjump_204_250}), we called it as a potential shock front. 
Other two equatorial shock fronts have significant discontinuities in both the temperature and surface brightness profiles (see Figures \ref{fig:NTjump_50_105} \& \ref{fig:STjump_235_300}).

\subsection{Optical, Radio and X-ray connection} 
Figure \ref{fig:Xsb_R600} shows  610 MHz GMRT radio contours
overlaid on the X-ray surface brightness map. The diffuse radio emission has three-component, e.g., a radio halo at the center of the cluster, a radio phoenix at the southeast to the center of the cluster, and a Head-Tail (HT) galaxy towards the southwest of the cluster center. The distribution of diffuse radio emission is not identical to the X-ray emission of the cluster; however, the elongation of the radio phoenix and the radio halo is similar to the merging axis of the cluster (see Figure \ref{fig:merging_scenario}).
Interestingly, the SE shock found at the place of the radio phoenix reported in \citep{Mandal_a2019A&A...622A..22M}. At the position of the SE shock, there are two extended radio emission lobes of the phoenix on each side (north and south of the sock). We assume that these extended radio emissions may be because of radio plasma spread by the shock on both sides. 
The radio halo appears to be bounded by the N and the S shock front (Figure \ref{fig:Xsb_R600}).
In Figure \ref{fig:Optical}, the X-ray peak and BCG2 does not coincide
with the BCG2, separated by $\sim$80 kpc, which is the signature of a post-merger cluster. A previous study by \citet{Barrena2013MNRAS.430.3453B} showed that the cluster has a binomial distribution of galaxies with peaks at BCG1 and BCG2. They also showed that the member galaxies are aligned towards a direction connected by the BCG1 and BCG2. It was proposed that as A1914 is situated in a node of cosmic filament, it is accumulating mass through this axis \citep{Barrena2013MNRAS.430.3453B}.
However, in our study, imaging analysis (see Figure \ref{fig:X-Ray-SB}) suggests that there is another merging axis towards the NW-SE direction.
The orientation of the radio emission also suggests the same. 
Detection of two equatorial shocks and an axial shock also support that merging is ongoing towards the NW-SE direction.
We assume that as the cluster is situated at the place of the cosmic node, there are possibilities of multiple merging axes in A1914 (Figure \ref{fig:X-Ray-SB}).

\subsubsection{Merging scenario and the origin of the radio halo}
\label{halo_origin}
\begin{figure}
\centering
\includegraphics[width=\columnwidth]{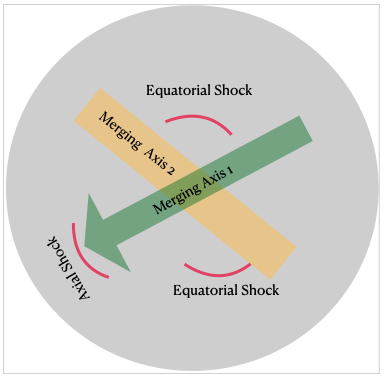}
\caption{This is the schematic diagram of the merging scenario in A1914. After visually analyzing maps in Figure \ref{fig:X-Ray-SB} and \ref{fig:Xsb_R600}, we draw this schematic diagram which explains the best plausible merging scenario in A1914. The merging axis 1 and 2 represent the axis found using X-ray and optical data analysis, respectively. (We refer readers to \citealt{Barrena2013MNRAS.430.3453B} for optical study and details of merging axis 2.)
}
\label{fig:merging_scenario}
\end{figure}

The cluster X-ray surface brightness and temperature distribution are complex and disturbed. The central part is elongated in the north-west/south-east direction and making $\sim 28\deg$ angle with respect to the E-W direction. 
\citet{Botteon_2018MNRAS.476.5591B} suggested a bullet-like merger in A1914.
Maps from imaging analysis shown in Figure \ref{fig:X-Ray-SB} have multiple signatures (shock fronts and morphology) of a bullet-like cluster merger.
To describe the merging process in A1914, we assume a general overview of binary mergers similar (maybe with some different initial conditions) to the case described in \citet{Ha2018ApJ...857...26H} and \citet{van_Weeren2019SSRv..215...16V}. As two mass clumps approached together and compressed, shock waves form and move outwards in the equatorial plane, perpendicular to the merger axis, called equatorial shocks. Later, two axial shocks launch along the merger axis \citet{Ha2018ApJ...857...26H}. This scenario is illustrated in Figure \ref{fig:merging_scenario}. We may not be able to detect one of the axial shocks due to the projection effect.
The merger scenario in A1914 may be much more complex.
The binary merger with nonzero impact factor can also create complex ICM distribution in the cluster environment, similar to what we observed in A1914 (please see the Galaxy Cluster Merger Catalog\footnote{http://gcmc.hub.yt/}). 
\citet{Barrena2013MNRAS.430.3453B} also reported another merging axis (see Figure \ref{fig:merging_scenario}) towards the Northeast and Southwest direction using optical and dark matter simulation. There may be multiple minor mergers and secondary in-fall mass clumps along the connecting filaments as there is a multiple merging axis present in the cluster.
These multiple mergers may excite turbulent gas flow motions; consequently, the formation of merger shocks proceeds in a more complex way than in idealized binary mergers.

Merging shocks also generate cluster-wide turbulence, which may re-accelerate the relativistic electrons present within the ICM, resulting in a radio halo. This radio halo seems to be bounded by two shocks N and S (Figure \ref{fig:Xsb_R600}). There are other examples of radio halo edges coincident with X-ray shocks (e.g. in Coma cluster, \citealt{Brown2011MNRAS.412....2B}; Abell 520,
\citealt{Markevitch2005ApJ...627..733M}; Abell 665, \citealt{Govoni2004ApJ...605..695G,Feretti2004A&A...423..111F}; Bullet Cluster, \citealt{Markevitch2002,Shimwell2014MNRAS.440.2901S}).

\subsubsection{The SE shock and the Origin of the radio phoenix}
\label{phoenix_origin}
The density jump at the southeast edge is higher compared to the other two shock fronts (see Table \ref{tab:djump}).
However, from spectral analysis,  we found that the deprojected temperature around the southeast edge is $T_{down}$ = $8.6^{+1.46}_{-1.01}$ keV and $T_{up} = 6.2^{+1.61}_{-1.10}$ keV, respectively. The gradient of both the temperature and brightness profiles are in the same direction (Figure \ref{fig:SETjump_204_250}), which signifies for the shock front.
The  Mach number of the shock is found to be $M_T = 1.4^{+0.52}_{-0.4}$ and $M_{SB} = 1.28^{+0.04}_{-0.02}$ using temperature and surface brightness jumps respectively.
At the position of this shock front, a radio phoenix was reported by \citet{Mandal_a2019A&A...622A..22M}.
A three-dimensional magneto-hydrodynamical simulation shows that radio phoenixes are aged radio galaxy lobes whose emission is boosted by adiabatic compression by a merger shock wave \citep{Enblin2001A&A...366...26E,Enblin2002MNRAS.331.1011E}.
Radio phoenices produced by adiabatic compression have filamentary morphology and ultra-steep or curved radio spectra \citep{Enblin2002MNRAS.331.1011E}.
If we consider the diffusive shock acceleration (DSA) model for this diffuse radio emission,  this SE shock might be able to re-accelerate electrons via Fermi-type processes. In this case, the integrated synchrotron spectrum should have a flat spectrum.
For the DSA model, radio spectral index is connected with the shock Mach number via equation \ref{eq:alpha_dsa} \citep{Blandford_1987PhR...154....1B}.
\begin{equation} \label{eq:alpha_dsa}
    \alpha = \frac{M^2 + 1}{ M^2 - 1}
\end{equation} \label{eq:radio_mach}
Therefore, the expected spectral index is found to be very steep, $\alpha $ = $2.6^{+1.49}_{-0.46}$, which is consistent with the spectral index $\alpha = 2.17 \pm 0.11$ reported in \citet{Mandal_a2019A&A...622A..22M}.
At the position of the SE shock, there are filamentary extensions on both sides (towards the north and south of the SE shock). These extended filamentary features imply the DSA at the vicinity of the shock region.
However, \citet{Mandal_a2019A&A...622A..22M} also found 
hint of curvature in the integrated spectrum of the radio phoenix.
This ultra-steep spectrum (or presence of curvature) of the radio phoenix disfavor for the possibility of shock acceleration since, in the case of shock acceleration, the spectrum tends to get flatter \citep{Mandal_a2019A&A...622A..22M}. 
Their detailed radio study on this radio phoenix ruled out the possibility of DSA, and they support the adiabatic compression scenario.
Therefore, detection of SE shock front at the place of radio phoenix further strengthens the adiabatic compression scenario proposed by \citet{Mandal_a2019A&A...622A..22M}. However, the extended filamentary structure at the position of the SE shock also suggest for the DSA model for the radio emission.

\section{Conclusions}   \label{Conclusions}
We report the detection of three X-ray shocks based on the newly released archival (120 ks) \textit{Chandra} X-ray observation (total 146 ks).
The objectives of this study have been to identify the thermodynamic discontinuities present in the X-ray maps and investigate if they correlate with the diffuse radio emission present in the cluster A1914. We summarize the essential results derived from the current analysis:

\begin{enumerate}
    \item Radio emission of a cluster depends on the dynamical state of the cluster. Signature of merging activities can imprint on the X-ray surface brightness map. We use an unsharp masked surface brightness image, GGM filter image, an RGB image of the surface brightness map (see Figure \ref{fig:X-Ray-SB}) to investigate the dynamical state of the cluster. A schematic diagram of the merging scenario is illustrated in Figure \ref{fig:merging_scenario}. We predict that the cluster might be experiencing multiple mergers; a strong one is a bullet-like merger in the NW-SE direction. However, an off-axis binary cluster merger with a large impact parameter can also create similar morphology as of the A1914.
    
    \item The thermodynamic maps of the cluster can also reveal the dynamical state of the cluster. Here, we create temperature maps of this cluster using three different techniques (see Appendix for WVT and Condbin temperature maps). These temperature maps show disturbed morphology, which is the signature of the cluster merger.
    The temperature and entropy maps (see Figure \ref{fig:press_ent}) also show the disturbed structure, indicating merging activities in the cluster. 
    
    \item We detected {two equatorial shock fronts (N \& S shown in Figure
    \ref{fig:NTjump_50_105} and \ref{fig:STjump_235_300}), and one tentative axial shock (SE, Figure \ref{fig:SETjump_204_250}).} The mach numbers of the shocks are listed in Table \ref{tab:djump} and \ref{tab:Tjump_t}, calculated from density and temperature jump, respectively.
    We found no discontinuity at the west region where previously shock was reported.
    
    \item The SE shock was found at the place of radio phoenix, which reveals that the adiabatic compression scenario of fossil plasma is the most favorable scenario on the origin radio phoenix. Some of the features (e.g., a filamentary structure at the position of SE shock) also hint towards the DSA for the origin of this radio structure.
    
    \item As the cluster is merging, the possible cause for the origin of the radio halo might be due to the re-acceleration of relativistic electrons by the cluster-wide turbulence created by merging shock. We also found that the radio halo appears to be bounded by N \& S shocks.
\end{enumerate}

\section*{Acknowledgements}
We would like to thank the anonymous reviewer for suggestions and comments that have helped to improve this paper. MR would like to thank Manoneeta Chakraborty, Juhi Tiwari, and Arnab Chakraborty for fruitful discussions. 
MR acknowledges the support provided by the DST-Inspire fellowship program (IF160343) by DST, India. 
RR is supported through grant ECR/2017/001296 awarded to AD by DST-SERB, India. 
We thank IIT Indore for providing computing facilities for the data analysis. 
We also thank GMRT staff for making these radio observations possible. GMRT is run by the National Centre for Radio Astrophysics of the Tata Institute of Fundamental Research.
The scientific results from the X-ray observations reported in this article are based on data obtained from the \textit{Chandra} X-ray Observatory (CXO). 
This research has also made use of the software provided by the \textit{Chandra} X-ray Center (CXC) in the application packages CIAO, ChIPS, and Sherpa.
This research made use of APLpy \citep{aplpy2012,aplpy2019}, an open-source plotting package for Python.

\section*{DATA AVAILABILITY}
The X-ray observations underlying this article are available in \textit{Chandra} data archive at [https://cda.harvard.edu/chaser/mainEntry.do], and can be accessed with Obs-IDs 542, 3593, 18252, 20023, 20024, 20025, and 20026. Radio data are available in GMRT archive, which can be accessed using ID 31\_049. The derived data presented in this article will be shared on reasonable request to the corresponding author.



\bibliographystyle{mnras}
\bibliography{reference}
\appendix
\section{Other X-ray Temperature Map Making Methods}

\subsubsection{Weighted Voronoi Tessellation (WVT) Method} \label{WVT}
This method was developed by \citet{Diehl2006} and widely used by X-ray astronomers to derive temperature map of the cluster of galaxies (e.g., A4059 \citep{Mernier2015}, 1RXS J0603.3+4214 \citep{Ogrean2013}, A85 \citep{Schenck2014}, A3667 \citep{Datta2014}, NGC 1404 \citep{Su2017}, Fornax \citep{Su2017b}).
The WVT binning algorithm creates non-overlapping regions based on the threshold SNR. SNR is calculated using merged CLEAN data and CLEAN background. Here we have used SNR 60, which comprises 116 WVT regions. 
After creating regions, we used \textit{dmextract} build-in \textit{CIAO} task to extract spectra from both source and background for each observation separately for the same region. Weighted response matrix (RMF) and weighted effective area (ARF) are calculated using \textit{specextract} for each observation. 
Spectral analysis is the same as discussed for the ACB method.

\subsubsection{Contour Binning (Contbin) Method}  \label{Contbin}
In this method, all the spectral regions were created by using an algorithm developed by \citet{Sanders2006}. These regions are defined based on the contours of constant X-ray SB to satisfy the SNR criteria. It starts at the highest flux pixel on a smoothed image, adds pixels nearest in SB to a bin until the SNR exceeds a threshold. A new region is then created.
SNR 60 has been used, which comprises 91 regions.
Spectral extraction and analysis were performed similarly to that of ACB and WVT methods.

\begin{figure*}
\centering
\begin{tabular}{lccr}
\includegraphics[width=\columnwidth]{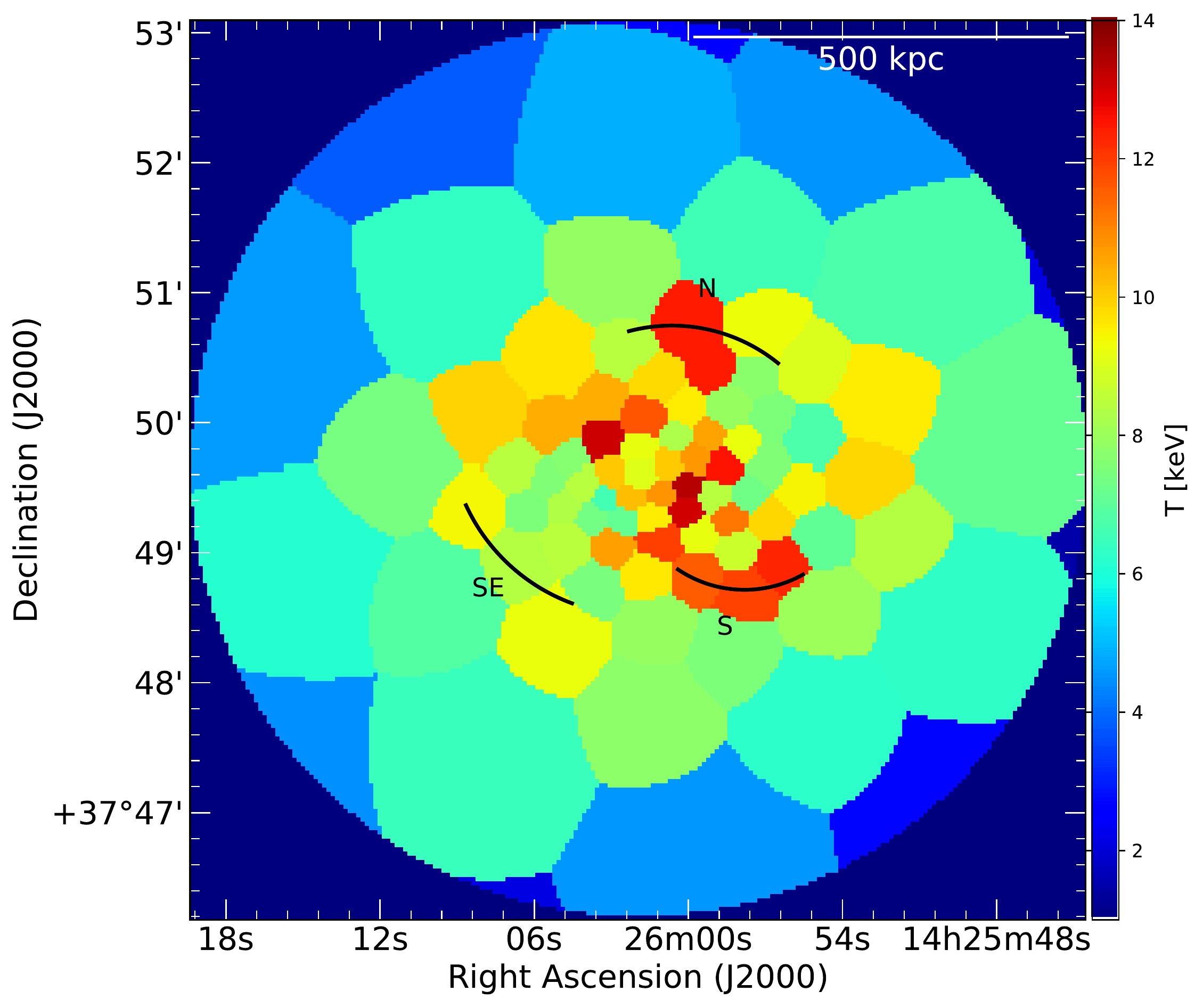} &
\includegraphics[width=\columnwidth]{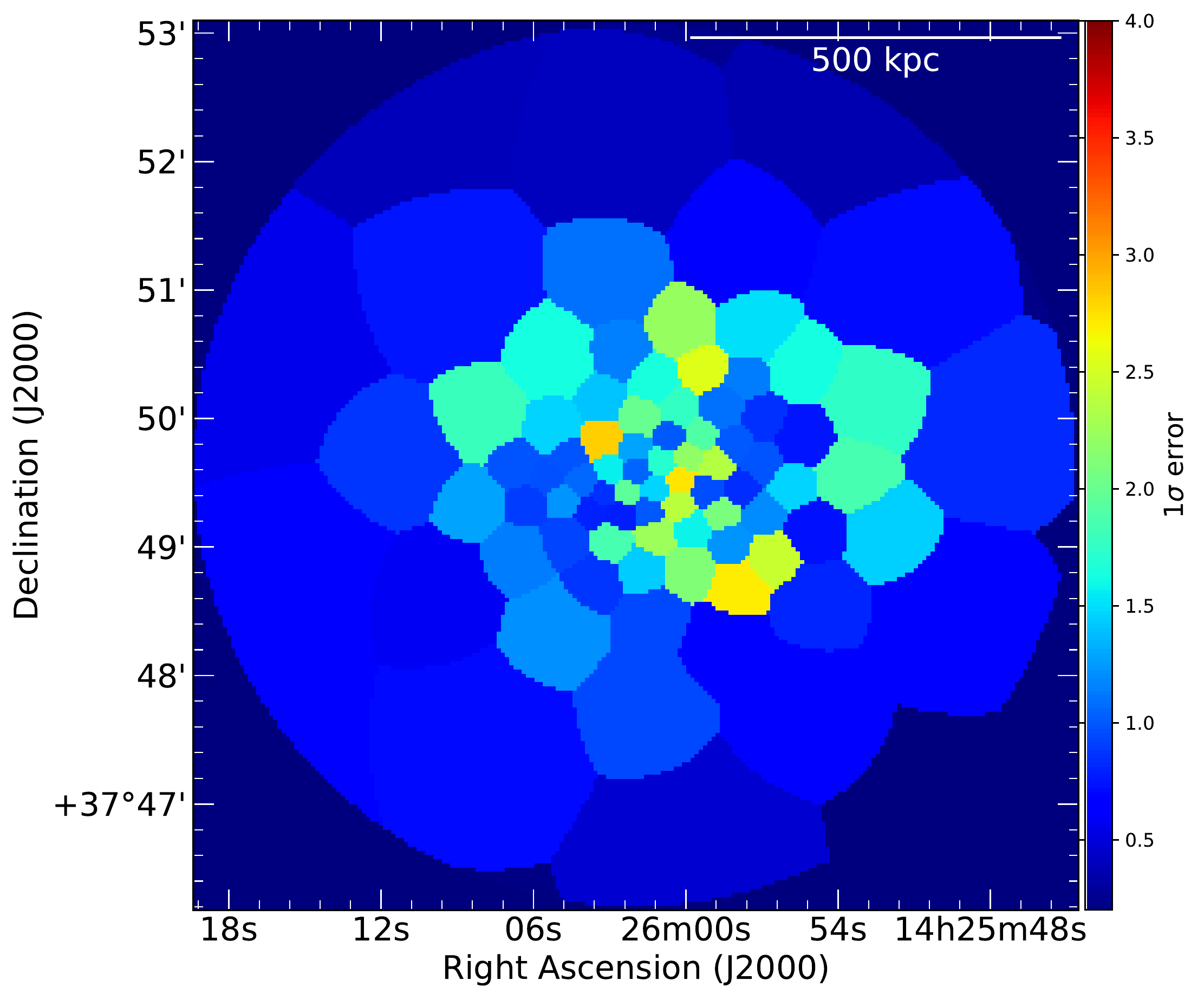} \\
\includegraphics[width=\columnwidth]{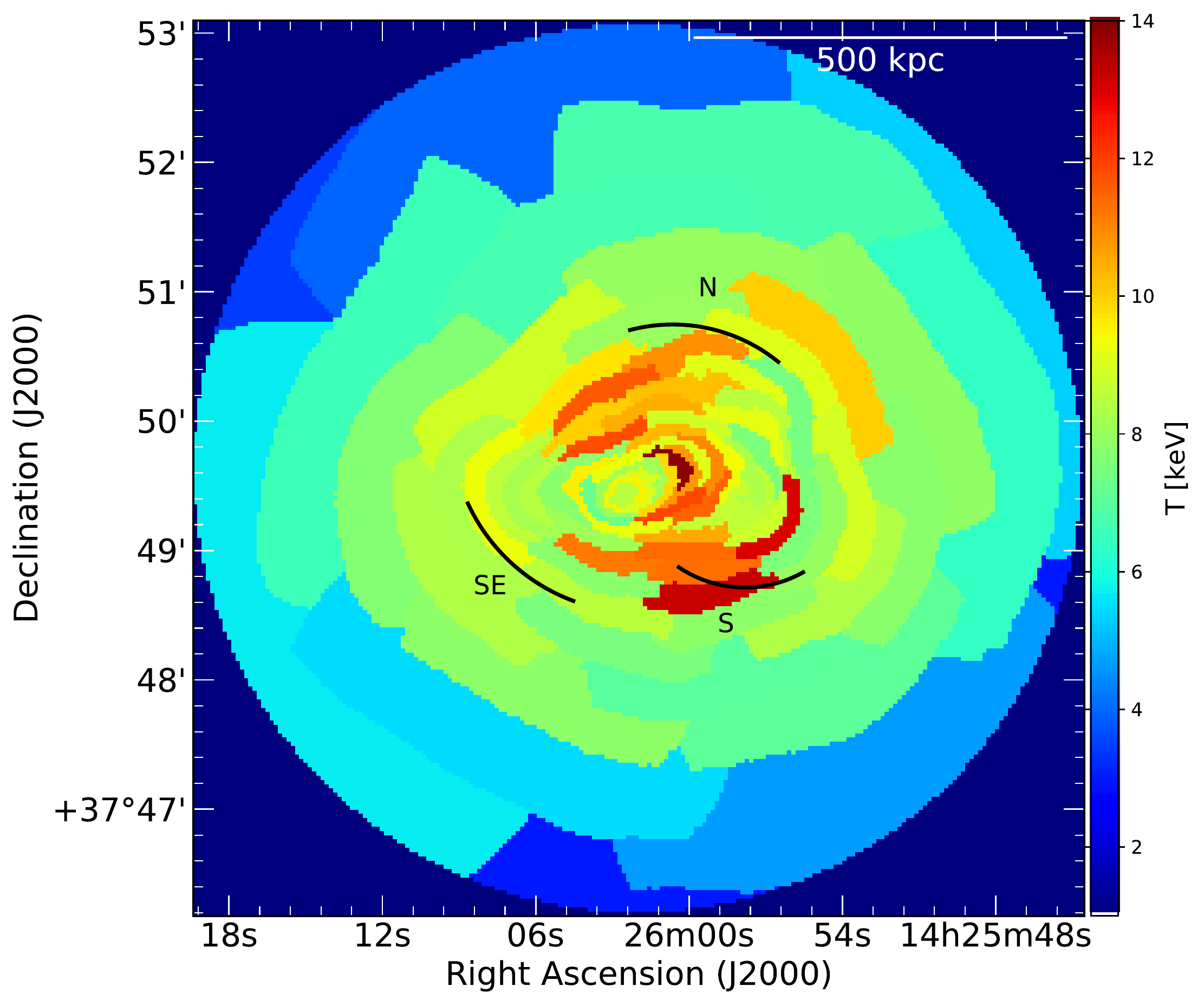} &
\includegraphics[width=\columnwidth]{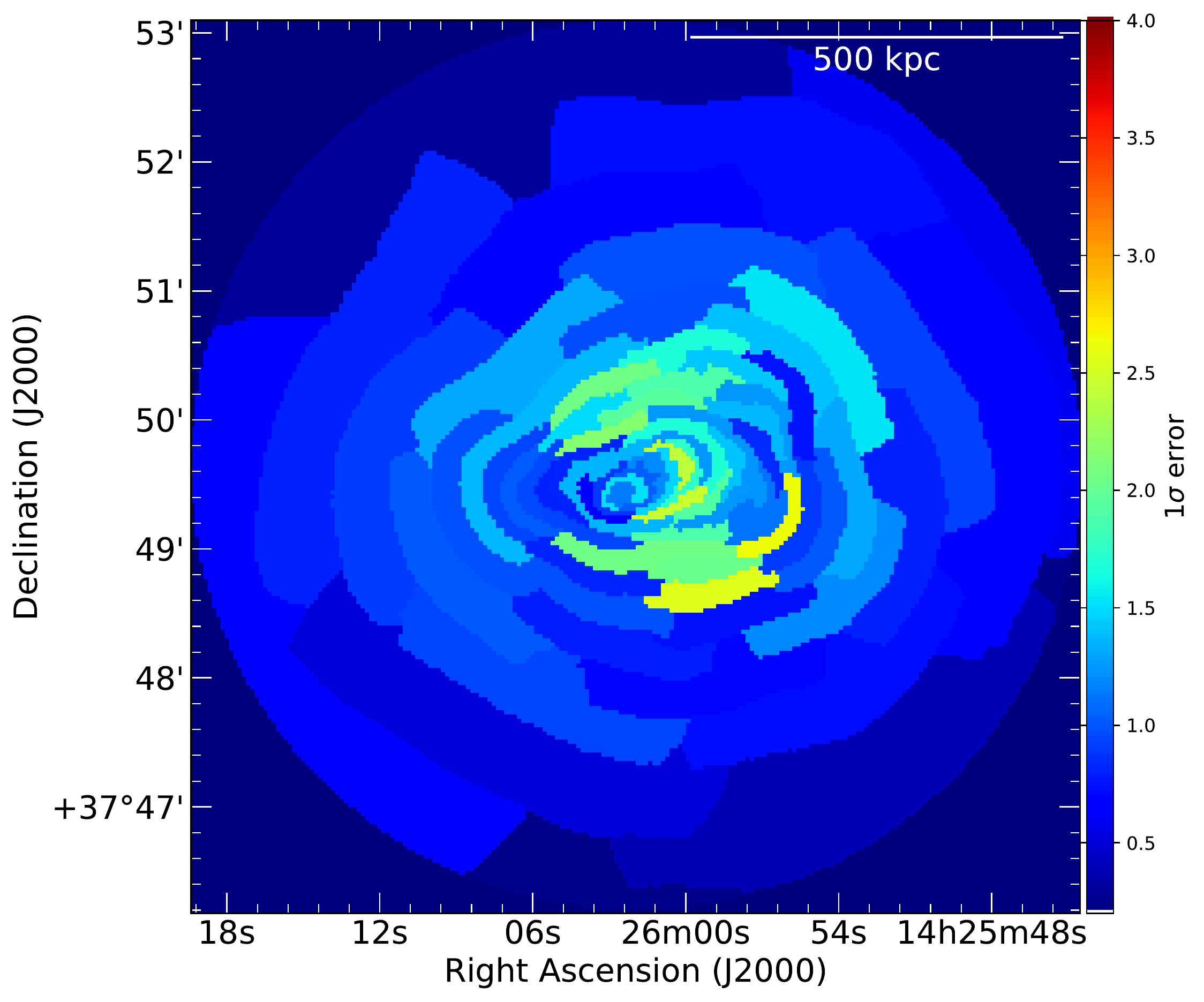} \\
\end{tabular}
\caption{\textit{Chandra} X-ray ACB temperature maps of the A1914 using other two different techniques, WVT temperature map (top left), and Contbin temperature map (bottom left).  Due to the low counts within the outer region of the cluster, we were only allowed to make a temperature map of the inner 600kpc regions similar to ACB temperature map in Figure \ref{fig:ACB_Tmap}.
Fractional error maps within 1$\sigma$ error are presented at the right panel of the corresponding temperature maps. All these temperature maps show a similar disturbing temperature structure all over the cluster, which reveals that the cluster is undergoing a merging event. To further investigate the significance of these temperature fluctuation, we have plotted temperature profile (\ref{fig:NTjump_50_105}, \ref{fig:STjump_235_300}, and Figure \ref{fig:SETjump_204_250}) along the high temperature regions (wedge are shown in top left panel of Figure \ref{fig:X-Ray-SB}) using classical fitting process. Arcs (N, S, and SE in black) represent the position of the shock waves.
These structures also resemble the RGB image of the X-ray surface brightness map presented in the bottom left, Figure \ref{fig:X-Ray-SB}.}
\label{fig:wvt_contbin_Tmap}
\end{figure*}


\bsp	
\label{lastpage}
\end{document}